\documentclass[reprint,a4paper]{article}
\usepackage{eurosym}
\usepackage[affil-it]{authblk}
\usepackage{setspace}
\usepackage{color}
\usepackage[dvipsnames]{xcolor}
\usepackage{amsmath}
\usepackage{amsfonts}
\usepackage{subcaption}
\usepackage[title]{appendix}
\usepackage{comment}
\usepackage{booktabs}
\usepackage{verbatim}
\usepackage{amssymb}
\usepackage{graphicx}
\usepackage{hyperref}
\hypersetup{
    colorlinks=true,
    linkcolor=blue,
    filecolor=magenta,
    urlcolor=[rgb]{0.00,0.00,0.50},
    citecolor=red,
    }
\usepackage[nameinlink]{cleveref}
\usepackage{cite}
\usepackage{bbold}
\usepackage{dsfont}
\usepackage{url}
\usepackage{caption}
\usepackage{subcaption}
\usepackage[top=2cm, bottom=2cm, left=3cm, right=3cm]{geometry}
\usepackage{upgreek}
\usepackage{physics}

\newcommand{\RE}{\operatorname{Re}}
\newcommand{\IM}{\operatorname{Im}}

\definecolor{darkgreen}{rgb}{0,0.35,0}

\newcommand{\CECs}{Centro de Estudios Cient\'{\i}ficos (CECs), Casilla 1469, Valdivia, Chile}
\newcommand{\USS}{Universidad San Sebasti\'{a}n, sede Valdivia, General Lagos 1163, Valdivia 5110693, Chile}
\newcommand{\UACh}{Instituto de Ciencias F\'isicas y Matem\'aticas, Universidad Austral de Chile, Casilla 567, 5090000 Valdivia, Chile}
\newcommand{\UCharles}{IPNP - Faculty of Mathematics and Physics, Charles University, V Hole\v{s}ovi\v{c}k\'ach 2, 18000 Prague 8, Czech Republic}
\begin{document}

\title{Fractional vorticity, Bogomol'nyi-Prasad-Sommerfield systems and complex structures for the (generalized)
spinor Gross-Pitaevskii equations}
\author[1,2]{Fabrizio Canfora\thanks{fabrizio.canfora@uss.cl}}
\author[3,4]{Pablo Pais\thanks{pais@ipnp.troja.mff.cuni.cz}}

\affil[1]{\USS} \affil[2]{\CECs} \affil[3]{\UACh} \affil[4]{\UCharles}

\date{}
\maketitle

\begin{abstract}
The (generalized) Gross-Pitaevskii equation (GPE) for a complex scalar
field in two spatial dimensions is analyzed. It is shown that there is
an infinite family of self-interaction potentials which admit Bogomol'nyi-Prasad-Sommerfield
(BPS) bounds together with the corresponding first-order BPS systems. For
each member of this family, the solutions of the first-order BPS systems
are automatically solutions of the corresponding second-order generalized
GPE. The simplest topologically non-trivial solutions of these first-order
BPS systems describe configurations with quantized fractional vorticity.
The corresponding fraction is related to the degree of non-linearity. The
case in which the self-interaction potential is of order six (namely
$|\Psi |^{6}$, which is a relevant theory both in relativistic quantum
field theories in $(2+1)$ dimensions in connection with the quantum Hall
effect as well as in the theory of the supersolids) is analyzed in detail.
Such formalism can also be extended to the case of quantum mixtures with
multi-component GPEs. The relationship between these techniques and supersymmetry
will be discussed. In particular, despite several common features, we will
show that there are multi-component GPEs that are not supersymmetric (at
least, not in the standard sense) and possess a BPS system of the above
type.
\end{abstract}

\newpage


\section{Introduction}

\label{sec-1}

Many of the most relevant open problems in physics are non-perturbative
in nature. For instance, the QCD phase diagram (especially at low temperatures
and finite density) is determined mainly by topological solitons such as
instantons, calorons, non-Abelian monopoles and so on (see
\cite{R0,R11,R2,Pisarski1,bibVTEX,bibVTEXa,bibVTEXb,WeinbergBook} and references therein).
The same is true for superconductors, where superconducting vortices with
quantized magnetic flux play a fundamental role. As far as the topological
solitons mentioned here above, a fundamental tool to study them is the
availability of Bogomolnyi-Prasad-Sommerfield (BPS) bounds for the free
energy of the configurations in terms of the relevant topological charges
(which can be the instanton number, the non-Abelian magnetic charge, the
magnetic flux and so on). The configurations which saturate these bounds
are called BPS solitons and are especially important as these minimize
the (free) energy in any given topological sector and, consequently, are
topologically stable. Moreover, the conditions to saturate the BPS bound
are first-order non-linear differential equations, which are considerably
easier to solve than the second-order field equations (one can also show
that when the BPS equations are satisfied, the second-order field equations
are satisfied too). The BPS equations allow both to analyze the low-energy
dynamics of these non-perturbative configurations using the so-called moduli-space
approximation (see \cite{modulia,modulib,moduli2} and references therein)
as well as the interactions with fermions through many powerful index theorems
\cite{WeinbergBook}. Consequently, the BPS equations are not important
simply because the field equations become easier to solve at the BPS points
of parameters space of the theory of interest (indeed, in the case of vortices
in superconductors at critical coupling, the BPS equation cannot be solved
analytically). In fact, BPS systems are especially relevant because very
powerful packages of exact non-perturbative results become available. Such
packages provide explicit descriptions of static and dynamical features
of these BPS solitons as well as the corresponding physical effects on
fermions, which would be very difficult to achieve for generic non-BPS
configurations.

There is another area where quantized vortices play a fundamental role
but where (until very recently) no BPS equations were available: namely,
superfluids and the corresponding vortices described by the Gross-Pitaevskii
equation (GPE)
\cite{sign1,sign2,sign3m,GPE2,GPE3,GPE4,GPE5,1suso,2suso,3suso,4suso,5suso,6suso,7susoa,7susob,7susoc,8suso,9suso,10suso,11suso,12suso,13suso,14suso}. It was usually assumed that the GPE in two or more spatial dimensions
does not possess a first-order BPS system with the sought property that,
when the BPS system is satisfied, the GPE is automatically satisfied. The
reason for this pessimistic view lies in the obvious BPS bound where, on
the right-hand side of the bound, the vorticity appears and cannot be saturated.

Using the techniques\footnote{%
Such techniques provided the first explicit examples (in the gauged non-linear
sigma model minimally coupled with electromagnetism) of BPS bounds in which
the topological charge appearing on the right-hand side of the bound is
a non-trivial function of the ``obvious topological charge''.} introduced
in \cite{BPSlast1,BPSlast2,BPSlast3}, it has been possible to construct
(for the first time in two spatial dimensions, to the best of authors knowledge)
a BPS system and the corresponding BPS bound for the GPE. As it is natural
to expect, the solutions of the first-order BPS system are solutions of
the second-order GPE automatically, with the usual quartic non-linear interaction
$|\Psi |^{4}$. Such BPS system possesses many non-trivial properties (which
will be analyzed in the following sections). Here however it is worth mentioning
that the topological charge appearing on the right hand side of the BPS
bound is a non-trivial combination of the vorticity (which is the ``obvious
topological charge'' in this context) with the topological charge
$Q_{2}$ which has been defined in \cite{BPSlast3}.

There are many situations (both in non-relativistic and relativistic settings)
where the relevant non-linear interaction term is expected to be different.
For instance, in the Tonks-Girardeau limit (see \cite{CS4a,CS4b} and references
therein), the non-linear interaction potential appearing in the corresponding
GPE is $|\Psi |^{6}$. In a Bose-Einstein condensate (BEC), such non-linear
interactions of order six in the GPE are related to atomic three-body processes,
and there are many situations in which such processes cannot be neglected.
Moreover, in relativistic field theories in (2+1) dimensions, the most
general polynomial interaction which is renormalizable at order six
\cite{Peskin}. Such a sixth-order interaction plays a fundamental role
in the theory of Abelian Chern-Simons vortices and its connections to the
quantum Hall effect (QHE); see \cite{CS1,CS2,CS3,sextic0,sextic3} and references therein. The
$\Psi ^{6}$ self-interaction potential is closely related to the unique
features of physics in $(2+1)$ dimensions as, for instance, particles may
possess fractional statistics as well as fractional spin with a behavior
which is somehow in between bosons and fermions (this is the so-called
anyonic behavior typical of the fractional quantum Hall effect: see
\cite{CS1,CS2,CS3} and references therein). A Chern-Simons term in the action
is able to take these remarkable effects into account. Vorticity (being
the natural topological charge in this context) plays a fundamental role
in this respect, and the self-interaction potential of the corresponding
complex scalar field can be a non-linear interaction of order 6 (see
\cite{sextic1} and references therein). In conclusion, many reasons (both
from relativistic and non-relativistic viewpoints) make a complex scalar
field with $|\Psi |^{6}$ interaction a particularly relevant theory. Quite
interestingly, (in the case in which the complex-scalar field is minimally
coupled to an Abelian Chern-Simons theory) a BPS bound is known (see
\cite{sextic0,sextic0.1,sextic0.2,sextic0.3,sextic3} and references therein)
while in the case of a complex scalar field with $|\Psi |^{6}$ alone no
BPS bound was known: we will fill this gap in the following sections.

Non-linear terms in the GPE, which are different from $|\Psi |^{4}$, are
also very relevant for the analysis of supersolid (a state of matter predicted
more than half a century ago: see \cite{1suso,2suso} and references therein).
In the supersolid phase, the system manifests at the same time two features
that (at a first glance) could appear quite the opposite, namely the crystalline
order and superfluidity. This means that some part of the total mass behaves
as a crystal, keeping another part alive that flows without dissipation.
This fact is one of the key signatures of supersolidity. In recent experiments,
much evidence supporting this phenomenon can be found (despite the first
difficulties \cite{3suso,4suso}): see
\cite{5suso,6suso,7susoa,7susob,7susoc,8suso} and references therein. In
this phase, quantum fluctuations play a fundamental role in stabilizing
the system (see review articles \cite{9suso,10suso,11suso}). In particular,
in the local density approximation (described by the Lee-Huang-Yang (LHY)
theory \cite{12suso,13suso,14suso}) the $\Psi ^{5}$, encodes very well
this ``stabilizing effects''. The present approach can also be extended
to such potential.

The considerations here above suggest the following natural questions:
\emph{Are suitable BPS bounds (related to the vorticity of the complex scalar
field) still available even when different non-linear terms are included
in the GPE? How does the value of the critical rotation, which allows the
existence of regular vortices, depend on the non-linear interaction potential?
Can this formalism be extended to quantum mixtures satisfying a multi-component
version of the GPE?}

All these important questions will be answered in the case of
$|\Psi |^{5}$ and $|\Psi |^{6}$ interactions using the hidden complex
structure available in field space. One of the outcomes of the present
formalism is that the fraction which defines the fractional vorticity of
the simplest topologically non-trivial solutions of these BPS systems is
related to the degree of non-linearity of the self-interaction potential
of the generalized GPE.

Such a complex structure also allows the generalization of the present
results to the case of quantum mixtures described by the spinor multi-component
GPE. The generalization of the multi-component GPE discloses a very important
feature of present formalism. The complex structure in field space, which
is responsible for the appearance of these BPS systems associated with
relevant (generalized) GPE, shares several features with supersymmetric
(SUSY) scalar field theories (see \cite{SUSY} and references therein).
However, we will construct interesting examples of spinor GPE that are
not supersymmetric (at least, not in the usual sense) and possess an associated
BPS system despite the lack of SUSY.

This paper is organized as follows: in Section~\ref{sec-2}, we present
the BPS system and the simplest case of GPEs. In Section~\ref{sec-3}, the
fractional-vorticity solutions for fifth- and sixth-order GPEs are shown.
The case of multi-component GPEs is revised in Section~\ref{sec-4} under
our method. Section~\ref{sec-5} is devoted to the conclusions and final
remarks. We will provide the computational details in three appendices.


\section{GPE, complex structure and fractional vorticity}

\label{sec-2} 

Let us consider the following first-order elliptic system in two spatial
dimensions (where
$\Phi _{1}=\Phi _{1}\left ( \overrightarrow{x}\right ) $,
$%
\Phi _{2}=\Phi _{2}\left ( \overrightarrow{x}\right ) $ are the two unknown
real scalar fields while $\overrightarrow{x}=(x,y)$ are two-dimensional
Cartesian coordinates)
%
\begin{eqnarray}
\partial _{x}\Phi _{1}+\partial _{y}\Phi _{2} &=&A\left ( \Phi _{1},
\Phi _{2}\right ) \ ,
\label{BPSs1}
\\
\partial _{y}\Phi _{1}-\partial _{x}\Phi _{2} &=&B\left ( \Phi _{1},
\Phi _{2}\right ) \ ,
\label{BPSs2}
\end{eqnarray}%
where $A\left ( \Phi _{1},\Phi _{2}\right ) $ and
$B\left ( \Phi _{1},\Phi _{2}\right ) $ are suitable non-linear functions
of the fields whose form is fixed, as it will be now shown, by the requirement
that the two unknown real scalar fields (or the unknown complex scalar
field) satisfy ``reasonable'' second-order field equations.

From the above first-order system, one can deduce that
$\Phi _{1}=\Phi _{1}\left ( \overrightarrow{x}\right ) $,
$\Phi _{2}=\Phi _{2}\left ( \overrightarrow{x}\right ) $ satisfy the following
second-order semilinear system of equations:
%
\begin{eqnarray}
\triangle \Phi _{1} &=&C_{1}\partial _{x}\Phi _{2}+C_{2}\partial _{y}
\Phi _{2}+\frac{1}{2}\frac{\partial }{\partial \Phi _{1}}\left ( A^{2}+B^{2}
\right ) \ ,
\label{RD1}
\\
\triangle \Phi _{2} &=&-C_{1}\,\partial _{x}\Phi _{1}-C_{2}\,
\partial _{y}\Phi _{1}+\frac{1}{2}
\frac{\partial }{\partial \Phi _{2}}\left ( A^{2}+B^{2}\right ) \ ,
\label{RD2}
\end{eqnarray}%
where%
%
\begin{equation}
C_{1}=\frac{\partial A}{\partial \Phi _{2}}+
\frac{\partial B}{\partial \Phi _{1}}\ ,\ \ C_{2}=
\frac{\partial B}{\partial \Phi _{2}}-
\frac{\partial A}{%
\partial \Phi _{1}}\ .
\label{RD3}
\end{equation}%
Thus, if one requires that in the system of coupled non-linear PDEs in
Eqs.~(\ref{RD1}) and (\ref{RD2}) first derivatives of the field should not appear
(as it is the case for the GPE), then $A$ and $B$ must be chosen as conjugated
Harmonic functions. Indeed, $C_{1}=C_{2}=0$ are equivalent to fields-space
Cauchy-Riemann equations.

On the other hand, it is worth emphasizing that this condition\footnote{Namely,
that the first derivatives of the fields should not appear in the second
order field equations.} is not mandatory at all. In this manuscript, we
will impose it since we are interested in the standard GPE (where the first
derivatives of the condensates are absent). However, the possibility to
include first derivatives is very interesting and can be useful in other
contexts (such as hydrodynamics, where both first and second derivatives
of the velocity field appear together in the corresponding field equations).
Hence, the techniques presented here can be adapted to far more general
situations: we will come back to this interesting issue in a future publication.

Therefore, when $C_{1}=C_{2}=0$, one gets
%
\begin{eqnarray}
\triangle \Phi _{1} &=&\frac{1}{2}
\frac{\partial }{\partial \Phi _{1}}\left ( A^{2}+B^{2}\right ) \ ,
\label{RD1.1}
\\
\triangle \Phi _{2} &=&\frac{1}{2}
\frac{\partial }{\partial \Phi _{2}}\left ( A^{2}+B^{2}\right ) \ .
\label{RD2.2}
\end{eqnarray}%
In other words, one can define a ``superpotential'' $W$, which is an analytic function of $%
Z=\Phi _{1}+i\Phi _{2}$ and then $A$ and $B$ are the real and imaginary
parts of $W$:
%
\begin{eqnarray}
W &=&W(Z),\ \ Z=\Phi _{1}+i\Phi _{2}\ ,
\label{superpotential1}
\\
A &=&\RE W\ ,\ \ B=\IM W\ .
\label{superpotential2}
\end{eqnarray}%
With the above choice, Eq.~(\ref{RD3}) is identically satisfied.

\subsection{The case of GPE with $|\Psi |^{4}$ interactions}
\label{sec2.1}

In a recent work \cite{BPSlast3}, using an educated guess, the first example
of a first-order BPS system for the GPE in two spatial dimensions (when
the GPE is not integrable) was discovered. In fact, the origin of this
first-order BPS system discussed in \cite{BPSlast3} can be easily understood
in terms of the present formalism encoded in Eqs.~(\ref{BPSs1}), (\ref{BPSs2}), (\ref{superpotential1}) and (\ref{superpotential2}). On the other hand,
the present formalism is far more powerful than the ``educated guess''
since it allows quite striking generalizations, as discussed in the following
sections.

If one considers the quadratic superpotential
%
\begin{equation}
W = \frac{\kappa}{\sqrt{2}}\, Z^{2}\ ,
\label{usualGPE}
\end{equation}%
one arrives at the following BPS system
%
\begin{eqnarray}
\frac{\partial \,\Phi _{1}}{\partial x}+
\frac{\partial \,\Phi _{2}}{\partial y} &=&\frac{\kappa}{\sqrt{2}}\,(
\Phi _{1}^{2}-\Phi _{2}^{2})\;,
\label{fo_GP_1}
\\
\frac{\partial \,\Phi _{1}}{\partial y}-
\frac{\partial \,\Phi _{2}}{\partial x} &=&\sqrt{2}\,\kappa \, \Phi _{1}
\,\Phi _{2}\ ,
\label{fo_GP_2}
\end{eqnarray}%
whose solutions are automatically solutions of the GPE
%
\begin{eqnarray}
-\triangle \Phi _{j}+g_{eff}\left ( \overrightarrow{\Phi }\cdot
\overrightarrow{\Phi }\right ) \Phi _{j} &=&0\ ,\ \ j=1,2
\label{GPE2.2}
\\
\left ( \Phi _{1}\right ) ^{2}+\left ( \Phi _{2}\right ) ^{2} &=&
\overrightarrow{%
\Phi }\cdot \overrightarrow{\Phi }\;,
\notag
\\
g_{eff}&=&\kappa ^{2} \;,
\notag
\end{eqnarray}%
while the energy becomes%
%
\begin{equation}
E=\frac{\hslash ^{2}}{2M}\int _{\Omega }d^{2}x\left [ \sum _{j=1}^{2}
\left ( \overrightarrow{\nabla }\Phi _{j}\right ) ^{2}+
\frac{g_{eff}}{2}\left ( \overrightarrow{\Phi }\cdot
\overrightarrow{\Phi }\right ) ^{2}\right ] \ ,
\label{GPE2.3}
\end{equation}%
\begin{equation*}
\left ( \overrightarrow{\nabla }\Phi _{j}\right ) ^{2}=\left (
\partial _{x}\Phi _{j}\right ) ^{2}+\left ( \partial _{y}\Phi _{j}
\right ) ^{2}\ .
\end{equation*}%
The mapping of the field variables $\Phi _{j}$, $j=1,2$ in terms of the
GPE wave function $\Psi $ is the following%
%
\begin{equation}
\Phi _{1}=\rho \cos S\ ,\ \Phi _{2}=\rho \sin S\ \Leftrightarrow \
\Phi _{1}=\RE \Psi \ ,\ \Phi _{2}=\IM \Psi \ ,\ \rho \geq 0\ ,\
\label{GPE2.1}
\end{equation}%
where $\rho $ is the amplitude of the wave function while $S$ is the phase.

The solutions of this first-order BPS system in Eqs.~(\ref{fo_GP_1}) and
(\ref{fo_GP_2}) possess quite remarkable properties.
\begin{enumerate}
\item \textbf{BPS bound}

It is a direct computation to show that the energy can be rewritten as
follows:
%
\begin{equation}
E=\frac{\hslash ^{2}}{2M}\int _{\Gamma }d^{2}x\left [\left (\partial _{x}
\Phi _{1}+\partial _{y}\Phi _{2}-A\right ) ^{2}+\left ( \partial _{y}
\Phi _{1}-\partial _{x}\Phi _{2}-B\right ) ^{2}\right ] +Q_{1}+Q_{2}
\;,
\label{BPS1-16}
\end{equation}%
where
%
\begin{equation}
A=\frac{\kappa}{\sqrt{2}}\left ( \Phi _{1}^{2}-\Phi _{2}^{2}\right )
\ \ ,\ \
B=\sqrt{2}\,\kappa \,\Phi _{1}\,\Phi _{2}\;,\ \ \kappa ^{2}=g_{eff}>0
\;.
\label{BPS2-17}
\end{equation}%
The topological charge is a combination of the following two boundary terms:
%
\begin{equation}
Q_{1}=\frac{\hslash ^{2}}{M}\int _{\Gamma }d^{2}x\Lambda \quad
\text{and}%
\quad Q_{2}=\frac{\hbar ^{2}\,\kappa }{\sqrt{2}\,M}\int _{\Gamma }d^{2}x
\left ( \partial _{x}J^{x}+\partial _{y}J^{y}\right ) \;,
\label{BPS3-18}
\end{equation}%
with
%
\begin{eqnarray}
\Lambda &=&\partial _{x}(2\Phi _{1}\partial _{y}\Phi _{2})+\partial _{y}(-2
\Phi _{1}\partial _{x}\Phi _{2})=d\Phi _{1}\wedge d\Phi _{2}=d\rho
\wedge dS=d\omega \;,
\notag
\label{BPS4-19}
\\
\omega &=&\rho dS\;,
\end{eqnarray}%
and
%
\begin{equation}
J^{x}=\Phi _{1}\left ( \frac{\Phi _{1}^{2}}{3}-\Phi _{2}^{2}\right )
\;,\quad J^{y}=\Phi _{2}\left ( \Phi _{1}^{2}-\frac{\Phi _{2}^{2}}{3}
\right ) \;.
\label{BPS5-20}
\end{equation}%

\item \textbf{$1/3$-fractional-vorticity configurations}

Let us consider the polar representation of the complex scalar field
\begin{eqnarray*}
\Phi _{1}(r,\theta ) &=&\rho (r,\theta )\cos \,S(r,\theta )\;,
\\
\Phi _{2}(r,\theta ) &=&\rho (r,\theta )\sin \,S(r,\theta )\;.
\end{eqnarray*}%
in cylindrical coordinates
\begin{eqnarray*}
r &=&\sqrt{x^{2}+y^{2}}\;,
\\
\theta &=&\arctan (\frac{y}{x})\;.
\end{eqnarray*}%
Then, taking $\frac{\partial \,S}{\partial \,\theta }=\frac{1}{3}$, we
arrive at the following analytic solution of the BPS system
\begin{equation*}
\rho (r)=\frac{2\sqrt{2}}{2\sqrt{2}\,A\,r^{\frac{1}{3}}-3\kappa \,r}
\;,
\end{equation*}%
and
%
\begin{eqnarray}
\Phi _{1}(r,\theta ) &=&
\frac{2\sqrt{2}\,\cos \,\left ( \frac{\theta }{3}%
+\theta _{0}\right ) }{2\sqrt{2}\,A\,r^{\frac{1}{3}}-\,3\kappa \,r}\;,
\label{solGPE1-21}
\\
\Phi _{2}(r,\theta ) &=&
\frac{2\sqrt{2}\,\sin \,\left ( \frac{\theta }{3}%
+\theta _{0}\right ) }{2\sqrt{2}\,A\,r^{\frac{1}{3}}-\,3\kappa \,r}\;.
\notag
\end{eqnarray}%
The above configuration is defined within annulus-like regions (where the
denominator does not vanish) and possesses fractional vorticity.\footnote{The
vorticity is computed by taking the inner radius of the annulus as the
short distance cut-off a, which is of the order of the size of the atoms
of the condensates (this is common practice also in the analysis of usual
superfluid vortices: see \cite{sign1,sign2,sign3m} and references therein).
Thus, only the outer boundary of the annulus contributes to the vorticity.
Thus, only the outer boundary of the annulus contributes to the vorticity.
In fact, in the present case, the topological charges appearing on the
right-hand side of the BPS bound play a more fundamental role.}
\end{enumerate}

\section{GPE with $|\Psi |^{5}$ and $|\Psi |^{6}$ interactions}
\label{sec-3}

Two natural questions arise: how much do these features depend on the non-linear
term? Can these techniques be adapted to more general self-interacting
complex scalar fields? We will analyze in detail two cases which are very
important both in the theory of GPE and in relativistic QFT in (2+1) dimensions
(in connection with Abelian Chern-Simons theory coupled to a relativistic
scalar field: see \cite{sextic1} and references therein).

\subsection{The case of the $|\Psi |^{5}$ interactions}
\label{sec3.1}

The theory of supersolids initiated long ago (see \cite{1suso,2suso} and
references therein). Recent experiments confirmed the appearance of this
state of matter (see \cite{5suso,6suso,7susoa,7susob,7susoc,8suso} and
references therein). In the local density approximation (see
\cite{12suso,13suso,14suso} and references therein) the $|\Psi |^{5}$ encodes
the stabilizing property needed to support supersolidity. Thus, we will
analyze the GPE with the $|\Psi |^{5}$ interaction.

Let us consider the following superpotential
%
\begin{eqnarray}
W &=&\frac{\kappa }{\sqrt{5/2}}\,Z^{\frac{5}{2}}=
\frac{\kappa }{\sqrt{5/2}}\,%
\left [ \Phi _{1}+\mathrm{i}\,\Phi _{2}\right ] ^{\frac{5}{2}}\ ,
\label{GPEsuso1}
\\
Z &=&\rho \,e^{\mathrm{i}\,S}\;\rightarrow \;Z^{\frac{5}{2}}=\rho ^{
\frac{5}{%
2}}\,e^{\mathrm{i}\,\frac{5}{2}\,S}\;,
\notag
\end{eqnarray}%
By substituting these to (\ref{BPSs1})--(\ref{BPSs2}), one arrives at the
following BPS system
\begin{eqnarray*}
\frac{\partial \,\Phi _{1}}{\partial x}+
\frac{\partial \,\Phi _{2}}{\partial y} &=&\frac{\kappa }{\sqrt{5/2}}
\,\rho ^{\frac{5}{2}}\,\cos (\frac{5}{2}%
\,S )\;,
\\
\frac{\partial \,\Phi _{1}}{\partial y}-
\frac{\partial \,\Phi _{2}}{\partial x} &=&\frac{\kappa }{\sqrt{5/2}}
\,\rho ^{\frac{5}{2}}\,\sin (\frac{5}{2}%
\,S)\;,
\end{eqnarray*}%
whose solutions are automatically solutions of the GPE (see Appendix~\ref{appendix_BPS_as_GPE} for details)
%
\begin{eqnarray}
-\triangle \Phi _{j}+g_{eff}\left ( \overrightarrow{\Phi }\cdot
\overrightarrow{\Phi }\right ) ^{\frac{3}{2}}\Phi _{j} &=&0\ ,\ \ j=1,2
\label{GPEsuso4}
\\
\left ( \Phi _{1}\right ) ^{2}+\left ( \Phi _{2}\right ) ^{2} &=&
\overrightarrow{%
\Phi }\cdot \overrightarrow{\Phi }\ .
\notag
\end{eqnarray}

The simplest topologically non-trivial solution of the BPS system takes
the form (see Appendix~\ref{appendix_fractional_vortex_solution})
%
\begin{eqnarray}
\label{sol_n_52}
\rho (r) &=&(A\,r^{3/7}-\frac{21}{4\sqrt{10}}\,\kappa \,r)^{-
\frac{2}{3}}\;,
\\
S (\theta ) &=&\frac{2}{7}\,\theta +S_{0}\;.
\notag
\end{eqnarray}%
Several comments are in order. First of all, the solution is well-defined
within an annulus. The extension of the annulus is bounded by the requirement
that $\rho (r)$ in the above equation must be well-defined. Secondly, vorticity
has a fractional value of $2/7$ which is tied to the degree of non-linearity
of the self-interaction potential.

\subsection{The case of the $|\Psi |^{6}$ interactions}
\label{sec3.2}

Among the many non-linear terms that can appear in the GPE besides the
usual quartic interaction, the most interesting is the $|\Psi |^{6}$ interaction,
which corresponds to three-body interactions in a dilute Bose gas. In particular,
it is well known that in the Tonks-Girardeau limit (see
\cite{CS4a,CS4b} and references therein), the $|\Psi |^{6}$ self-interaction
potential (instead of $|\Psi |^{4}$) plays the main role in the corresponding
GPE. In the theory of BECs, such non-linear interactions of order six in
the GPE are related to atomic three-body processes, and there are many
situations in which such processes cannot be neglected. Moreover, a complex
scalar field with $|\Psi |^{6}$ interactions plays a fundamental role in
the analysis of Abelian Chern-Simons theory coupled with a relativistic
scalar field in relation with the FQHE. Such theory allows transition from
type I to type II vortices (see \cite{sextic1} and references therein).
For these reasons, it is natural to ask whether or not such a theory supports
topologically non-trivial configurations even when the Abelian Chern-Simons
gauge field is turned off. Last but not least, in the case of relativistic
quantum field theories in $(2+1)$ dimensions, a non-linear polynomial interaction
of order six is the highest possible interaction, which is still renormalizable.

If one considers the quadratic superpotential
%
\begin{equation}
W=\frac{\kappa}{\sqrt{3}} Z^{3}\ ,
\label{GPE6}
\end{equation}%
then one arrives at the following BPS system
%
\begin{eqnarray}
\frac{\partial \,\Phi _{1}}{\partial x}+
\frac{\partial \,\Phi _{2}}{\partial y} &=&\frac{\kappa }{\sqrt{3}}\,(
\Phi _{1}^{3}-3\Phi _{1}\Phi _{2}^{2})\;,
\label{BPS61}
\\
\frac{\partial \,\Phi _{1}}{\partial y}-
\frac{\partial \,\Phi _{2}}{\partial x} &=&\frac{\kappa }{\sqrt{3}}\,
\left ( 3\Phi _{1}^{2}\,\Phi _{2}-\Phi _{2}^{3}\right ) \;,
\label{BPS62}
\end{eqnarray}%
whose solutions are automatically solutions of the GPE
%
\begin{eqnarray}
-\triangle \Phi _{j}+g_{eff}\left ( \overrightarrow{\Phi }\cdot
\overrightarrow{\Phi }\right ) ^{2}\Phi _{j} &=&0\ ,\ \ j=1,2
\label{GPE61}
\\
\left ( \Phi _{1}\right ) ^{2}+\left ( \Phi _{2}\right ) ^{2} &=&
\overrightarrow{%
\Phi }\cdot \overrightarrow{\Phi }\ ,\ \ g_{eff}=\kappa
\notag
\end{eqnarray}%
while the energy becomes
%
\begin{equation}
E=\frac{\hslash ^{2}}{2M}\int _{\Omega }d^{2}x\left [ \sum _{j=1}^{2}
\left ( \overrightarrow{\nabla }\Phi _{j}\right ) ^{2}+
\frac{g_{eff}}{2}\left ( \overrightarrow{\Phi }\cdot
\overrightarrow{\Phi }\right ) ^{3}\right ] \ ,
\label{GPE63}
\end{equation}%
\begin{equation*}
\left ( \overrightarrow{\nabla }\Phi _{j}\right ) ^{2}=\left (
\partial _{x}\Phi _{j}\right ) ^{2}+\left ( \partial _{y}\Phi _{j}
\right ) ^{2}\ ,
\end{equation*}%
\begin{equation*}
\Phi _{1}=\rho \cos S\ ,\ \Phi _{2}=\rho \sin S\ \Leftrightarrow \
\Phi _{1}=\RE \Psi \ ,\ \Phi _{2}= \IM \Psi \ ,\ \ \Psi =\rho \exp iS
\ ,\ \rho \geq 0\ .\
\end{equation*}

The energy can be rewritten as follows
%
\begin{equation}
E=\frac{\hslash ^{2}}{2M}\int _{\Gamma }d^{2}x\left [ \left (
\partial _{x}\Phi _{1}+\partial _{y}\Phi _{2}-A\right ) ^{2}+\left (
\partial _{y}\Phi _{1}-\partial _{x}\Phi _{2}-B\right ) ^{2}\right ] +Q_{1}+Q_{2}
\;,
\label{BPS1}
\end{equation}
where $A$ and $B$ are defined as
%
\begin{equation}
A=\frac{\kappa }{\sqrt{3}}\,(\Phi _{1}^{3}-3\Phi _{1}\Phi _{2}^{2})
\ \ ,\ \
B=\frac{\kappa }{\sqrt{3}}\,\left ( 3\Phi _{1}^{2}\,\Phi _{2}-\Phi _{2}^{3}
\right )\;.
\label{BPS2}
\end{equation}%
The topological charge is a combination of the following two boundary terms:
%
\begin{equation}
Q_{1}=\frac{\hslash ^{2}}{M}\int _{\Gamma }d^{2}x\Lambda \quad
\text{and}%
\quad Q_{2}=\frac{\hbar ^{2} }{2 M}\int _{\Gamma }d^{2}x\left (
\partial _{x}J^{x}+\partial _{y}J^{y}\right ) \;,
\label{BPS3}
\end{equation}%
where we defined
%
\begin{eqnarray}
\Lambda &=&\partial _{x}(2\Phi _{1}\partial _{y}\Phi _{2})+\partial _{y}(-2
\Phi _{1}\partial _{x}\Phi _{2})=d\Phi _{1}\wedge d\Phi _{2}=d\rho
\wedge dS=d\omega \;,
\notag
\label{BPS4}
\\
\omega &=&\rho dS\;,
\end{eqnarray}%
and
%
\begin{equation}
J^{x}=\frac{\,\kappa}{2\,\sqrt{3}}\,\left (\Phi _{1}^{4}+\Phi _{2}^{4}-6
\,%
\Phi _{1}^{2}\,\Phi _{2}^{2}\right ) \;,\quad J^{y}=
\frac{\kappa}{2\,\sqrt{3}}%
\,\left (4\Phi _{1}^{3}\,\Phi _{2}-4\Phi _{2}^{3}\,\Phi _{3}\right )
\;.
\label{BPS5}
\end{equation}%
It is clear from Eqs.~(\ref{BPS4}) and (\ref{BPS5}) that $Q_{1}$ and
$Q_{2}$ are integrals of total derivatives.

\subsubsection{The BPS system in polar representation}
\label{sec3.2.1}

\begin{figure}[h]
\begin{subfigure}[b]{0.45\textwidth}
\includegraphics[width=\textwidth]{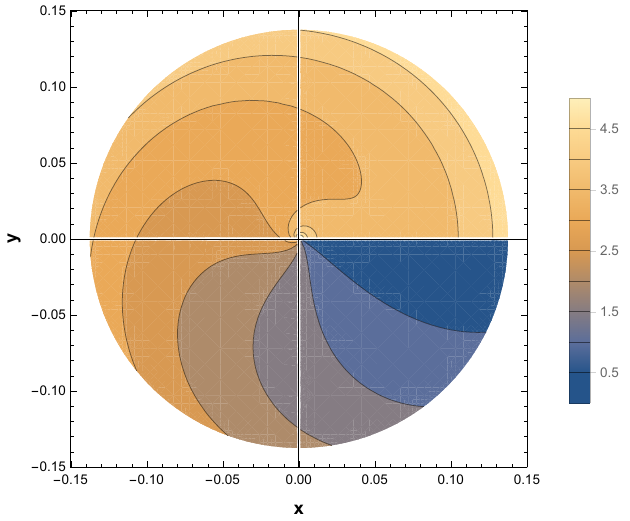}
\caption{Level curves for $\Phi_{1}$ of solution (\ref{solGPE1}). We see a branch cut in the positive $x$-axis due to the $\frac{\theta}{4}$ angular dependence of $\Phi_{1}$. }
\end{subfigure}
\hfill
\begin{subfigure}[b]{0.45\textwidth}
\includegraphics[width=\textwidth]{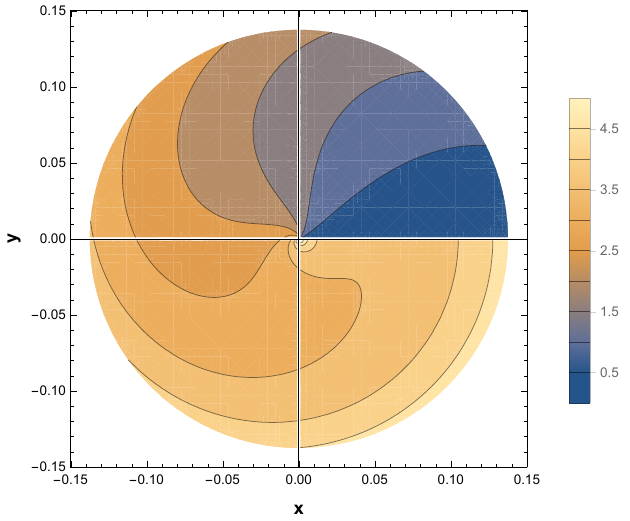}
\caption{Level curves for $\Phi_{2}$ of solution (\ref{solGPE1}). We see a branch cut in the positive $x$-axis due to the $\frac{\theta}{4}$ angular dependence of $\Phi_{2}$.}
\end{subfigure}
\caption{Level curves for solution (\protect\ref{solGPE1}). We took $\protect%
\kappa =1$ and $A=1$, and internal radius $a=0.3$ while the external one is $%
R\approx 0.91$ in arbitrary units. This solution can be embedded on a Riemann surface, as the fractional vorticity feature implies there should be a branch cut. In this case, such a branch cut lies on the positive $x$-axis, as is evident in the colour discontinuity in both real and imaginary parts of the solution.}
\label{fig_level_first_sol}
\end{figure}

As in the previous section, let us consider the polar representation for
the complex scalar field
%
\begin{eqnarray}
\Phi _{1}(r,\theta ) &=&\rho \,\cos (S)\;,
\notag
\label{solGPE1}
\\
\Phi _{2}(r,\theta ) &=&\rho \,\sin (S)\;.
\end{eqnarray}%
In terms of these variables, the simplest topologically non-trivial solution
of the BPS system for the $|\Psi |^{6}$ self-interaction potential takes
the form
%
\begin{eqnarray}
\rho (r) &=&(A\,r^{1/2}-\frac{4}{\sqrt{3}}\,\kappa \,r)^{-\frac{1}{2}}
\;,
\label{sol_n_3}
\\
S(\theta ) &=&\frac{\theta }{4}+S_{0}\;,
\notag
\end{eqnarray}%
where $A$ and $S_{0}$ are two integration constants. The extension of the
annulus is bounded by the requirement that $\rho (r)$ in the above equation
must be well-defined, and, as in the case of the $|\Psi |^{5}$ self-interaction potential, such a value is tied to the degree of non-linearity.
Fig.~\ref{fig_level_first_sol} shows the level curves for solution (\ref{solGPE1}).
There is a branch cut (which can be chosen on the positive $x$-axis), and
this is expected from the argument $\frac{\theta }{4} $ in (\ref{solGPE1}).
Hence, we can envisage this solution on a Riemann surface, similar to defects
with angular excess living on surfaces with negative intrinsic curvature
(see \cite{Kleinert} and references therein). These kinds of defects are
of paramount importance in using graphene as an effective playground to
test important features of the quantum field theories in curved spacetimes
\cite{Iorio2013,IorioReview,Iorio2018}.

On the other hand, it is worth emphasizing that a widespread and effective
technique to regularize the short-range divergences which appear in the
GPE is via a short-range cutoff (which corresponds to the atomic size of
the elementary constituents of the condensate itself). In particular, an
interesting question is whether or not multi-vortices solutions of the
BPS equations (\ref{BPS61}) and (\ref{BPS62}) can be constructed. The similarity
of these BPS equations with the ``zero-curvature condition'' for a complex
non-Abelian connection $\Xi $,
\begin{equation*}
d\Xi +\Xi \wedge \Xi =0\ ,
\end{equation*}%
where $\Xi $ is a one-form taking values in the Lie algebra of a suitable
Lie group, is apparent. In both cases, the linear terms contain derivatives
of the fields, while the non-linear terms are polynomial in the fields.
Thus, assuming that multi-vortices solutions can be constructed using slight
modifications of the standard techniques available in the presence of a
zero-curvature representation: see, for instance,
\cite{Hitchin1999,Babelon2003,Jurdjevic2016}, it will be possible to regularize
them by introducing the cutoff corresponding to the atomic size. In this
case, the regularization via atomic cutoff will play a prominent role (to
be explored in a future publication). On the other hand, in the case analyzed
in the present manuscript of a single defect with (fractional) vorticity,
both the regularization via atomic cutoff and ring-type domains work equally
well.

Let us stress also that configurations with fractional topological charges
are expected both in condensed matter physics (see
\cite{FracVort1,FracVort2,FracVort3,FracVort4} and references therein) and in
high-energy physics (see
\cite{FracVort5,FracVort6,FracVort7,FracVort8,FracVort9} and references therein).
On the other hand, usually, such fractional topological charges need some
extra ingredients to appear (as the above references clarify). In fact,
within the present framework,
\textit{the appearance of a fractional vorticity arises without any extra
ingredients.} In other words, it is just the non-linear term of the generalized
GPE which supports these configurations with fractional vorticity (the
corresponding fraction being fixed by the degree of non-linearity).

For the solution defined above, the two boundary terms $Q_{1}$ and
$Q_{2}$ contributing to the topological charge can be computed explicitly.
Taking into account Eq.~(\ref{BPS3}) for $S_{0}=0$, in an annulus, one
gets
%
\begin{equation}
\label{Q1_m_3}
Q_{1} =-\frac{\hbar ^{2}}{M}\int _{0}^{2\pi }\,\rho ^{2}\,\cos ^{2}S
\,\partial _{\theta }S\,d\theta = -
\frac{\pi \,\hbar ^{2}}{%
4\,M\,(A\,R^{1/2}-\frac{4}{\sqrt{3}}\,\kappa \,r)} +
\frac{\pi \,\hbar ^{2}}{%
4\,M\,(A\,a^{1/2}-\frac{4}{\sqrt{3}}\,\kappa \,r)} \;,
\end{equation}
where $R$ is the outer radius, and $a$ is the inner (which could be taken
as the size of the condensed atoms of interest).

Clearly, $Q_{1}$ is related to the vorticity as costume calculated. Indeed,
if we give boundary conditions at spatial infinity (so the radius
$R$ of the annulus $\Gamma $ approaches to infinity and the radius of the
atom $a$ to zero \cite{GPE5}) such that the amplitude $\rho $ goes to
$1$, then $Q_{1}$ becomes:
\begin{equation*}
\rho \underset{r\rightarrow \infty }{\rightarrow }1\ \ \text{and}\ \
\
\partial \Gamma =\text{``the circle at infinity''}\ \Rightarrow Q_{1}=-
\frac{%
\hslash ^{2}}{M}\,\frac{\partial \,S}{\partial \,\theta}\;.
\end{equation*}%
The same is true if one requires that $\rho =1$ on
$\partial \Gamma $ even when $\partial \Gamma $ is not the circle at infinity.
Therefore, the vorticity is just
$\frac{\partial \,S}{\partial \,\theta}$, which in this case has the fractional
value $\frac{1}{4}$.

As far as $Q_{2}$ is concerned, one gets at the same annulus
%
\begin{equation}
Q_{2}=\frac{\hslash ^{2}\,\kappa }{\sqrt{2}\,M}\oint _{\partial
\Gamma }\hat{n%
}\cdot \overrightarrow{J}\,ds\;,
\label{charge2}
\end{equation}%
where
\begin{eqnarray*}
J^{x} &=&\frac{\kappa}{2\,\sqrt{3}}\,\rho ^{4}\,\cos (4S)\;,
\\
J^{y} &=&\frac{\kappa}{2\,\sqrt{3}}\,\rho ^{4}\,\cos (4S)\;.
\end{eqnarray*}%
Hence, we have
\begin{eqnarray*}
\hat{n}\cdot \overrightarrow{J} &=&\cos \theta \, J^{x}+\sin \theta
\, J^{y}=%
\frac{\kappa}{2\,\sqrt{3}}\,\rho ^{4}\,\left ( \cos \theta \,
\cos (4\,S)+\sin \theta \,\sin (4\,S)\right )
\\
&=&\frac{\,\kappa}{2\,\sqrt{3}}\,\rho ^{4}\,\cos (4\,S-\theta )=
\frac{\,\kappa%
}{2\,\sqrt{3}}\,\rho ^{4}\;.
\end{eqnarray*}%
Therefore, we obtain
\begin{equation}
Q_{2} = \frac{\pi \,\hbar ^{2}\,\kappa}{{2\sqrt{3}M}}\int \limits _{a}^{R}%
\rho ^{4}(r)\,dr\, \;,
\notag
\label{Q2_first_sol}
\end{equation}
which can be computed using $\rho (r)$ in the solution (\ref{sol_n_3}).

One important property of this solution is the amplitude-phase relation
they should satisfy (see Appendix~\ref{appendix_amplitude_phase_relation} for a general relation). In this
case,
%
\begin{equation}
\Delta \,S=\sqrt{3}\,\kappa \,\rho ^{2}\,\left ( \cos (4\,S-\theta )
\,\frac{%
\partial \,S}{\partial r}+\frac{\sin (4\,S-\theta )}{r}\,
\frac{\partial \,S}{%
\partial \theta }\right ) \;.
\label{amp_pha_rel_n3}
\end{equation}%
One consequence of the above relation is that, excluding the case
$S(\theta )=\frac{\theta }{4}+2n\pi $, $n\in \mathbb{Z}$,
$S(\theta )$ is close to being harmonic (namely,
$\Delta \,S\approx 0$). Thus, in such a situation, $%
\rho $ must be close to zero\footnote{%
This is a reasonable conclusion as close to the position of any vortex
$%
S\sim \arctan \frac{y}{x}$ (where the vortex under investigation has been
taken as the origin of the coordinates system), and
$\arctan \frac{y}{x}$ is harmonic (excluding the origin itself). Hence,
close to the origin, $\rho $ must be small.}: $\rho \approx 0$. A significant
conclusion is that the BPS system provides a closed formula for the amplitude
$\rho $ in terms of the phase $S$. Such a formula is
\textit{valid for all solutions of the BPS equations and not only for the
simple solution found here}. Hence, the present approach reduces the unknown
functions of the GPE system from two to just one. Using the above expression
for $\rho $ in terms of the phase and its derivatives together with the
BPS system, one can derive a single master equation which only involves
$S$. Such a master equation can be used to construct multi-vortices configurations
numerically. We will return to analyzing this master equation in a future
publication.

\subsection{General considerations on the GPE}
\label{sec3.3}

In the previous sections, we have considered in detail only two cases (namely,
the GPE with $|\Psi |^{5}$ and $|\Psi |^{6}$ interactions) to show the
power of the present technique in the clearest possible way. However, it
is worth emphasizing that the superpotentials leading to these interactions
(as well as the $Z^{2}$ superpotential, which generates the simplest GPE
introduced in \cite{BPSlast3}) can, of course, be combined. For instance,
if as the superpotential one takes
\begin{equation*}
W=\kappa _{2}Z^{2}+\kappa _{5/2}Z^{5/2}+\kappa _{3}Z^{3}\ ,
\end{equation*}%
then one would get an extended version of the GPE equation with
$|\Psi |^{4}$%
, $|\Psi |^{5}$ and $|\Psi |^{6}$ interactions plus several cross-terms.
The applications of this superpotential will be discussed in the future
publications. In conclusion, the techniques introduced in this manuscript
allow us to detect the presence of BPS configurations even in many theories
(both relativistic and non-relativistic) of interest where no BPS bound
was previously known.

\section{Multicomponent GPEs: the case of two components}
\label{sec-4}

As it is well known, for weakly interacting mixtures of two BEC configurations,
a suitable theoretical tool is a generalization of the GPE where each condensate
is described by its corresponding wave function. Obviously, the two condensates
$\Psi _{1}$ and $\Psi _{2}$ (which are the two order parameters of the
two components of the mixture) interact, and this leads to a coupled system
of two GPEs (see
\cite{mixture1,sign1,sign2,sign3m,GPE2,GPE3,GPE4,GPE5,R0} and references therein).
The parameter which couples $\Psi _{1}$ and $\Psi _{2}$ (usually called
$g_{12}$) is determined by the scattering length of the process where an
atom in state 1 scatters from an atom in state 2. A further fundamental
effect in the analysis of mixtures is the possibility of transferring atoms
from one state to the other (this process can, for instance, be generated
by suitably tuned external fields). The theoretical description of this
process can be achieved by including a term proportional to
$\Psi _{2}$ in the GPE for $\Psi _{1}$ and \textit{vice-versa}. The typical
form of the stationary GPE for two interacting condensates is:%
%
\begin{eqnarray}
\frac{\hbar ^{2}\triangle }{2m}\Psi _{1} &=&\left ( V_{ext}+g_{11}
\left \vert \Psi _{1}\right \vert ^{2}+g_{12}\left \vert \Psi _{2}
\right \vert ^{2}\right ) \Psi _{1}-\frac{\hbar \Omega _{(1)}}{2}
\Psi _{2}
\label{mixture1}
\\
\frac{\hbar ^{2}\triangle }{2m}\Psi _{2} &=&\left ( V_{ext}+V_{hf}+g_{12}
\left \vert \Psi _{1}\right \vert ^{2}+g_{22}\left \vert \Psi _{2}
\right \vert ^{2}\right ) \Psi _{2}-\frac{\hbar \Omega _{(2)}}{2}
\Psi _{1}\ ,
\label{mixture2}
\end{eqnarray}%
where $V_{hf}$ is the hyperfine splitting, $g_{11}$ ad $g_{22}$ are the
coupling constants of $\Psi _{1}$ and $\Psi _{2}$ respectively and the
Rabi energies $\hbar \Omega _{(i)}$ ($i=1,2$) are responsible for the possibility
to transfer atoms from one state to the other.

The present formalism can be extended to the case of interacting quantum
mixtures. Here, for simplicity, it will be discussed the interesting limiting
case\footnote{%
However, as the following subsection will clarify, the tools introduced
in the present manuscript are far more general in applicability.} in which
$%
g_{11}$ and $g_{22}$ are negligible with respect to $g_{12}$ (however,
as it will be clear from the following sections, the present formalism
has a far wider range of applicability). In order to show the power of
the method, instead of assuming that the Rabi energies
$\hbar \Omega _{(i)}$ are just constants which do not depend on
$\Psi _{1}$ and $\Psi _{2}$, the following intriguing situation will be
discussed:
%
\begin{equation}
\hbar \Omega _{(1)}\approx \left \vert \Psi _{1}\right \vert ^{2}\ \
\ %
\text{and}\ \ \ \hbar \Omega _{(2)}\approx \left \vert \Psi _{2}
\right \vert ^{2}.
\label{mixture3}
\end{equation}%
Namely, the possibility to transfer atoms from $\Psi _{1}$ to
$\Psi _{2}$ is proportional to $|\Psi _{1}|^{2}$ and vice-versa. This is
a very reasonable hypothesis as, if $|\Psi _{1}|^{2}\approx 0$,
\textit{then there is no atom to transfer to begin with}. It is worth emphasizing
that such a generalization is challenging to study with the available analytic
tools, while the formalism proposed here allows to easily derive a BPS
first-order system even for Eqs.~(\ref{mixture1}), (\ref{mixture2}) and
(\ref{mixture3}).

Hence, let us consider two condensates $\Psi _{1}$ and $\Psi _{2}$ described
by the following\footnote{%
The system of coupled stationary GPEs in Eqs.~(\ref{mixFinal1}) and (\ref{mixFinal2})
is also extremely relevant to the relations of the present formalism with
SUSY are concerned (this issue will be discussed in the following subsections).}
coupled stationary GPEs
%
\begin{eqnarray}
-\triangle \Psi _{1}+g_{1}g_{2}\left \vert \Psi _{1}\right \vert ^{2}
\Psi _{2}+g_{1}^{2}\left \vert \Psi _{2}\right \vert ^{2}\Psi _{1} &=&0
\ ,
\label{mixFinal1}
\\
-\triangle \Psi _{2}+g_{1}g_{2}\left \vert \Psi _{2}\right \vert ^{2}
\Psi _{1}+g_{2}^{2}\left \vert \Psi _{1}\right \vert ^{2}\Psi _{2} &=&0
\ ,
\label{mixFinal2}
\\
\left ( \Phi _{1}\right ) ^{2}+\left ( \Phi _{2}\right ) ^{2} &=&
\left \vert \Psi _{1}\right \vert ^{2}\ ,
\notag
\\
\left ( \Phi _{3}\right ) ^{2}+\left ( \Phi _{4}\right ) ^{2} &=&
\left \vert \Psi _{2}\right \vert ^{2}\ .
\notag
\end{eqnarray}%
The interpretation of the various terms in the above system of equation
is the following. The term $g_{1}^{2}|\Psi _{2}|^{2}\Psi _{1}$ in Eq.~(\ref{mixFinal1})
and the term $g_{2}^{2}|\Psi _{1}|^{2}\Psi _{2}$ in Eq.~(\ref{mixFinal2})
represent the coefficients of the nonlinear interaction terms related to
the s-wave scattering length between the atoms of states $1$ and
$%
2$. The term $g_{1}\,g_{2}\,|\Psi _{1}|^{2}\,\Psi _{2}$ in Eq.~(\ref{mixFinal1})
and the term $g_{1}g_{2}|\Psi _{2}|^{2}\Psi _{1}$ in Eq.~(\ref{mixFinal2})
describe the transfer of atoms from one state to the other. On the other
hand, Eqs.~(\ref{mixFinal1}) and (\ref{mixFinal2}) also encode a very reasonable
phenomenological hypothesis: that the possibility to transfer atoms from
$\Psi _{1}$ to $\Psi _{2}$ is proportional to $|\Psi _{1}|^{2}$ and vice-versa.
Indeed, the larger is $|\Psi _{1}|^{2}$, the more likely is this transfer.
When $|\Psi _{1}|^{2}$ and $|\Psi _{2}|^{2}$ are approximately constant,
the terms $g_{1}\,g_{2}\,|\Psi _{1}|^{2}\,\Psi _{2}$ in Eq.~(\ref{mixFinal1})
and $g_{1}\,g_{2}\,|\Psi _{2}|^{2}\,\Psi _{1}$ in Eq.~(\ref{mixFinal2})
behave like the usual Rabi terms in Eqs.~(\ref{mixture1}) and (\ref{mixture2}).
It will be now discussed how one can construct a first-order BPS system
also for a coupled nonlinear system of GPEs such as Eqs.~(\ref{mixFinal1})
and (\ref{mixFinal2}).

The system of coupled GPEs in Eqs.~(\ref{mixFinal1}) and (\ref{mixFinal2})
possess the following first-order BPS system:
%
\begin{eqnarray}
\partial _{x}\Phi _{1}+\partial _{y}\Phi _{2} &=&g_{1}\left ( \Phi _{1}
\Phi _{3}-\Phi _{2}\Phi _{4}\right ) \ ,
\label{mixBPSF1}
\\
\partial _{y}\Phi _{1}-\partial _{x}\Phi _{2} &=&g_{1}\left ( \Phi _{1}
\Phi _{4}+\Phi _{2}\Phi _{3}\right ) \ ,
\label{mixBPSF2}
\\
\partial _{x}\Phi _{3}+\partial _{y}\Phi _{4} &=&g_{2}\left ( \Phi _{1}
\Phi _{3}-\Phi _{2}\Phi _{4}\right ) \ ,
\label{mixBPSF3}
\\
\partial _{y}\Phi _{3}-\partial _{x}\Phi _{4} &=&g_{2}\left ( \Phi _{1}
\Phi _{4}+\Phi _{2}\Phi _{3}\right ) \ ,
\label{mixBPSF4}
\end{eqnarray}%
\begin{equation*}
\Psi _{1}=\rho _{1}\,e^{\mathrm{i}\,S_{1}}\;,\quad \Psi _{2}=\rho _{2}
\,e^{%
\mathrm{i}\,S_{2}}\;.
\end{equation*}%
A direct computation reveals that any solution of Eqs.~(\ref{mixBPSF1})--(\ref{mixBPSF4})
is also a solution of Eqs.~(\ref{mixFinal1}) and (\ref{mixFinal2}). The system of Eqs.~(\ref{mixBPSF1})--(\ref{mixBPSF4}) can be interpreted
as a coupled first-order system for the BPS systems in the
$\Psi _{1}$ and $%
\Psi _{2}$ components. To have a first-order system describing interacting
condensates is a remarkable achievement as it brings in all the known results
on BPS solitons (such as the possibility to study solitons dynamics using
the moduli space approximation: see
\cite{modulia,modulib,moduli2,moduli3} and references therein). Moreover, being
a first-order system, it allows a direct hydrodynamics interpretation in
terms of the two superfluid components $\Psi _{1}$ and $\Psi _{2}$.

At this point, a natural question arises:
\textit{is it possible to determine whether or not the interaction terms
of a generic GPE system admit a first-order BPS system (corresponding to
suitable super-potentials)?} In the Ginzburg-Landau theory for BCS superconductors,
one can find a first-order BPS system that only implies the second-order
field equations at critical coupling. Regarding the non-Abelian monopoles
in the Georgi-Glashow model, one can construct a first-order BPS system,
which implies the second-order field equations only when the Higgs potential
vanishes. Thus, in general, one should expect that requiring that a given
GPE system of equations possesses a first order BPS system puts some restrictions
both on the coupling and on the form of the non-linear interactions. Although
it is generally difficult to answer this intriguing question, one can observe
what happens in the present case with two interacting condensates. The
generic GPE system for the situation described at the beginning of this
section is
%
\begin{eqnarray}
-\triangle \Psi _{1}+G_{1}\left \vert \Psi _{1}\right \vert ^{2}\Psi _{2}+G_{2}
\left \vert \Psi _{2}\right \vert ^{2}\Psi _{1} &=&0\ ,
\label{generic1}
\\
-\triangle \Psi _{2}+G_{3}\left \vert \Psi _{2}\right \vert ^{2}\Psi _{1}+G_{4}
\left \vert \Psi _{1}\right \vert ^{2}\Psi _{2} &=&0\ ,
\label{generic2}
\end{eqnarray}%
where the $G_{1}$, $G_{2}$, $G_{3}$ and $G_{4}$ are the four couplings
of the theory. The discussion above shows the requirement for the above
system in Eqs.~(\ref{generic1}) and (\ref{generic2}) possess a first-order
BPS system that can be satisfied provided
\begin{equation*}
G_{1}=g_{1}g_{2}\ ,\ \ G_{2}=g_{1}^{2}\ ,\ \ G_{3}=g_{1}g_{2}\ ,\
G_{4}=g_{2}^{2}\ ,
\end{equation*}%
(as one can easily see by comparing Eqs.~(\ref{generic1}) and (\ref{generic2})
with Eqs.~(\ref{mixFinal1}) and (\ref{mixFinal2})). Consequently, in order
for the system in Eqs.~(\ref{generic1}) and (\ref{generic2}) to possess
a first-order BPS system, the above restrictions on the coupling constants
$G_{1}$, $G_{2}$, $G_{3}$ and $G_{4}$ must be satisfied. We will come back
on this interesting issue in a future publication.

One can rewrite the first-order system (\ref{mixBPSF1})--(\ref{mixBPSF4})
by using the ``amplitude-phase'' representation; hence, in terms of the amplitude $\rho $ and the phase
$S$, the first-order BPS system in Eqs.~(\ref{mixBPSF1})--(\ref{mixBPSF4}),
after a change of variables $u_{j}=\ln \rho _{j}$, $j\in \{1,2\}$, and
some rearrangements, reduces to
\begin{eqnarray*}
\frac{\partial \,u_{1}}{\partial r}+\frac{1}{r}
\frac{\partial \,S_{1}}{%
\partial \theta } &=&g_{1}\,e^{u_{2}}\,\cos \,(\theta -2\,S_{1}-S_{2})
\;,
\\
\frac{\partial \,S_{1}}{\partial r}-\frac{1}{r}
\frac{\partial \,u_{1}}{%
\partial \theta } &=&g_{1}\,e^{u_{2}}\,\sin \,(\theta -2\,S_{1}-S_{2})
\;,
\\
\frac{\partial \,u_{2}}{\partial r}+\frac{1}{r}
\frac{\partial \,S_{2}}{%
\partial \theta } &=&g_{2}\,e^{u_{1}}\,\cos \,(\theta -2\,S_{2}-S_{1})
\;,
\\
\frac{\partial \,S_{2}}{\partial r}-\frac{1}{r}
\frac{u_{2}}{\partial \theta } &=&g_{2}\,\,\,e^{u_{1}}\,\sin \,(
\theta -2\,S_{2}-S_{1})\;.
\end{eqnarray*}%
The above form is very convenient for the hydrodynamics interpretation.
In particular, let us introduce the superfluid velocity
$\overrightarrow{V_{j}}=%
\overrightarrow{\nabla }S_{j}$, $j\in \{1,2\}$. In terms of $u_{j}$ and
$%
\overrightarrow{V_{j}}$, the above BPS system can be written as
%
\begin{eqnarray}
\frac{\partial \,u_{1}}{\partial r}+\frac{V_{1\,\theta }}{r} &=&g_{1}
\,e^{u_{2}}\,\cos \,(\theta -2\,S_{1}-S_{2})\;,
\label{hydrodynamical_sytstem}
\\
V_{1\,r}-\frac{1}{r}\frac{\partial \,u_{1}}{\partial \theta } &=&g_{1}
\,e^{u_{2}}\,\sin \,(\theta -2\,S_{1}-S_{2})\;,
\label{eq52}
\\
\frac{\partial \,u_{2}}{\partial r}+\frac{V_{2\,\theta }}{r} &=&g_{2}
\,e^{u_{1}}\,\cos \,(\theta -2\,S_{2}-S_{1})\;,
\label{eq53}
\\
V_{2\,r}-\frac{1}{r}\frac{u_{2}}{\partial \theta } &=&g_{2}\,\,\,e^{u_{1}}
\,%
\sin \,(\theta -2\,S_{2}-S_{1})\;,
\label{eq54}
\end{eqnarray}%
where $V_{j\,r}=\frac{\partial S_{j}}{\partial r}$ and
$V_{j\,\theta }=\frac{%
\partial S_{j}}{\partial \theta }$, $j\in \{1,2\}$.

One can see that the right-hand sides of the above BPS system couple the
two superfluid components. In particular, when
$\left \vert \theta -2\,S_{1}-S_{2}\right \vert $ is small, a
$\theta $-dependence of the amplitude $\rho _{1}$ induces a radial component
of the superfluid velocity. Moreover, in the regions where
$\left ( \theta -2\,S_{1}-S_{2}\right ) $ is constant, the first two BPS
equations reduce to an algebraic for $\rho _{1}$ and the velocity
$\overrightarrow{V}_{1}$ where $r$ and the coupling $\kappa $ appear as
parameters. Moreover, when $\left ( \theta -2\,S_{1}-S_{2}\right ) $ is
approximately constant, regions where $V_{1r}$ is not vanishing appear.
In these regions, $V_{1r}$ is tied to
$\frac{%
\partial \,\rho _{1}}{\partial \theta }$ (similar arguments hold in the
analysis of the third and fourth BPS equations). Thus, one can see that
the interplay between the two phases, $S_{1}$ and $S_{2}$, plays a fundamental
role which can be made explicit thanks to the BPS system. Such a hydrodynamics
description of the BPS system
\emph{allows us to algebraically express the superfluid velocities of the
two components in terms of the two amplitudes, their derivatives, and the
two phases}. This is quite a powerful simplification, that can facilitate
the numerical resolution of very complicated equations, which can only
be achieved thanks to the present approach (since the GPE and its spinor
generalization always involve both the velocities and the corresponding
derivatives). Even more, the system of equations (\ref{hydrodynamical_sytstem}),
being algebraic in $V_{j}$ can be used to visualize the behavior of a quantum
mixture of two or more components through direct plots $V_{j}$ vs. amplitude
$u_{j}$. We will come back to the hydrodynamics analysis of this first-order
BPS system in a future publication.

\subsection{The complex structure behind the BPS equation for multicomponent
superfluids}
\label{sec4.1}

In order to disclose the mathematical structure behind the (at first glance)
mysterious fact that any solution of Eqs.~(\ref{mixBPSF1})--(\ref{mixBPSF4})
is also a solution of Eqs.~(\ref{mixFinal1}) and (\ref{mixFinal2}), let
us consider the following a more general first-order elliptic system in
two spatial dimensions (where
$\Phi _{j}=\Phi _{j}\left ( \overrightarrow{x}%
\right ) $, $j\in \{1\,,2\,,3\,,4\}$ are the four unknown real scalar fields
while $\overrightarrow{x}=(x,y)$ are two-dimensional Cartesian coordinates)
%
\begin{eqnarray}
\partial _{x}\Phi _{1}+\partial _{y}\Phi _{2} &=&A\left ( \Phi _{1},
\Phi _{2},\Phi _{3},\Phi _{4}\right ) \ ,
\label{mix1}
\\
\partial _{y}\Phi _{1}-\partial _{x}\Phi _{2} &=&B\left ( \Phi _{1},
\Phi _{2},\Phi _{3},\Phi _{4}\right ) \ ,
\label{mix2}
\\
\partial _{x}\Phi _{3}+\partial _{y}\Phi _{4} &=&C\left ( \Phi _{1},
\Phi _{2},\Phi _{3},\Phi _{4}\right ) \ ,
\label{mix3}
\\
\partial _{y}\Phi _{3}-\partial _{x}\Phi _{4} &=&D\left ( \Phi _{1},
\Phi _{2},\Phi _{3},\Phi _{4}\right ) \ ,
\label{mix14}
\end{eqnarray}

From the above first-order system, one can easily deduce that
$\Phi _{j}=\Phi _{j}\left ( \overrightarrow{x}\right ) $, with
$j=1,2,3,4$ satisfy the following second-order semilinear system of elliptic
partial differential equations (without first derivatives of the
$\Phi _{j}\left ( \overrightarrow{x}\right ) $):
%
\begin{eqnarray}
\triangle \Phi _{1} &=&\frac{1}{2}
\frac{\partial }{\partial \Phi _{1}}\left ( A^{2}+B^{2}\right ) +C
\frac{\partial A}{\partial \Phi _{3}}+D
\frac{\partial B%
}{\partial \Phi _{3}}\ ,
\label{mix5}
\\
\triangle \Phi _{2} &=&\frac{1}{2}
\frac{\partial }{\partial \Phi _{2}}\left ( A^{2}+B^{2}\right ) +C
\frac{\partial A}{\partial \Phi _{4}}+D
\frac{\partial B%
}{\partial \Phi _{4}}\ ,
\label{mix6}
\\
\triangle \Phi _{3} &=&\frac{1}{2}
\frac{\partial }{\partial \Phi _{3}}\left ( C^{2}+D^{2}\right ) +A
\frac{\partial C}{\partial \Phi _{1}}+B
\frac{\partial D%
}{\partial \Phi _{1}}\ ,
\label{mix7}
\\
\triangle \Phi _{4} &=&\frac{1}{2}
\frac{\partial }{\partial \Phi _{4}}\left ( C^{2}+D^{2}\right ) +A
\frac{\partial C}{\partial \Phi _{2}}+B
\frac{\partial D%
}{\partial \Phi _{2}}\ ,
\label{mix8}
\end{eqnarray}%
\textit{provided}
%
\begin{eqnarray}
0 &=&\frac{\partial A}{\partial \Phi _{2}}+
\frac{\partial B}{\partial \Phi _{1}}\ ,\ \ 0=
\frac{\partial B}{\partial \Phi _{2}}-
\frac{\partial A}{%
\partial \Phi _{1}}\ ,
\label{cr1}
\\
0 &=&\frac{\partial A}{\partial \Phi _{4}}+
\frac{\partial B}{\partial \Phi _{3}}\ ,\ \ 0=
\frac{\partial B}{\partial \Phi _{4}}-
\frac{\partial A}{%
\partial \Phi _{3}}\ ,
\label{cr2}
\end{eqnarray}%
%
\begin{eqnarray}
0 &=&\frac{\partial C}{\partial \Phi _{4}}+
\frac{\partial D}{\partial \Phi _{3}}\ ,\ \ 0=
\frac{\partial D}{\partial \Phi _{4}}-
\frac{\partial C}{%
\partial \Phi _{3}}\ ,
\label{cr3}
\\
0 &=&\frac{\partial C}{\partial \Phi _{2}}+
\frac{\partial D}{\partial \Phi _{1}}\ ,\ \ 0=
\frac{\partial D}{\partial \Phi _{2}}-
\frac{\partial C}{%
\partial \Phi _{1}}\ ,
\label{cr4}
\end{eqnarray}%
Thus, $A$ and $B$ are conjugated Harmonic functions with respect to the
complex variable $Z_{1}=\Phi _{1}+i\Phi _{2}$ as well as with respect to
the complex variable $Z_{2}=\Phi _{3}+i\Phi _{4}$ (and the same is true
for $C$ and $D$).

In other words, one can define two ``superpotentials'' $W_{1}$ and
$W_{2}$ which are analytic functions both of
$Z_{1}=\Phi _{1}+i\Phi _{2}$ and of $%
Z_{2}=\Phi _{3}+i\Phi _{4}$. In this way, $A$ and $B$ are the real and
imaginary parts of $W_{1}$ while $C$ and $D$ are the real and imaginary
parts of $W_{2}$:
%
\begin{eqnarray}
W_{1} &=&W_{1}(Z_{1},Z_{2}),\ \ Z_{1}=\Phi _{1}+i\Phi _{2}\ ,\ Z_{2}=
\Phi _{3}+i\Phi _{4}
\label{superp1}
\\
A &=&\RE W_{1}\ ,\ \ B=\IM W_{1}\ ,
\label{superp2}
\\
W_{2} &=&W_{2}(Z_{1},Z_{2}),\ \ C=\RE W_{2}\ ,\ \ D=\IM W_{2}\ ,
\label{superp3}
\end{eqnarray}%
With the above choice, Eqs.~(\ref{cr1})--(\ref{cr4}) are identically satisfied.
It is worth emphasizing that the generalization of this formalism to three
or more components is straightforward.

The simplest example of mixture of two condensates possessing a first-order
BPS system corresponds to the choice
%
\begin{eqnarray}
W_{1} &=&g_{1}Z_{1}Z_{2},\ W_{2}=g_{2}Z_{1}Z_{2}
\label{mixBPS1}
\\
A &=&g_{1}\left ( \Phi _{1}\Phi _{3}-\Phi _{2}\Phi _{4}\right ) \ ,\
\
B=g_{1}\left ( \Phi _{1}\Phi _{4}+\Phi _{2}\Phi _{3}\right ) \ ,
\label{mixBPS2}
\\
C &=&g_{2}\left ( \Phi _{1}\Phi _{3}-\Phi _{2}\Phi _{4}\right ) \ ,\
\
D=g_{2}\left ( \Phi _{1}\Phi _{4}+\Phi _{2}\Phi _{3}\right ) \ ,
\label{mixBPS3}
\end{eqnarray}%
where $g_{1}$ and $g_{2}$ are the strength of the interactions. With the
above choice, the second-order field equations satisfied by the fields
are:
%
\begin{eqnarray}
\triangle \Phi _{1} &=&g_{1}g_{2}\left ( \Phi _{1}^{2}+\Phi _{2}^{2}
\right ) \Phi _{3}+g_{1}^{2}\left ( \Phi _{3}^{2}+\Phi _{4}^{2}
\right ) \Phi _{1}\ ,
\label{sofe1}
\\
\triangle \Phi _{2} &=&g_{1}g_{2}\left ( \Phi _{1}^{2}+\Phi _{2}^{2}
\right ) \Phi _{4}+g_{1}^{2}\left ( \Phi _{3}^{2}+\Phi _{4}^{2}
\right ) \Phi _{2}\ ,
\label{sofe2}
\\
\triangle \Phi _{3} &=&g_{2}^{2}\left ( \Phi _{1}^{2}+\Phi _{2}^{2}
\right ) \Phi _{3}+g_{1}g_{2}\left ( \Phi _{3}^{2}+\Phi _{4}^{2}
\right ) \Phi _{1}\ ,
\label{sofe3}
\\
\triangle \Phi _{4} &=&g_{2}^{2}\left ( \Phi _{1}^{2}+\Phi _{2}^{2}
\right ) \Phi _{4}+g_{1}g_{2}\left ( \Phi _{3}^{2}+\Phi _{4}^{2}
\right ) \Phi _{2}\ .
\label{sofe4}
\end{eqnarray}%
The above system of equations can be interpreted in terms of the mixture
of condensates as follows:%
\begin{eqnarray*}
\Phi _{1} &=&\rho _{1}\cos S_{1}\ ,\ \Phi _{2}=\rho _{1}\sin S_{1}\
\Leftrightarrow \ \Phi _{1}=\RE \Psi _{1}\ ,\ \Phi _{2}=\IM \Psi _{1}
\ ,\ \
\Psi _{1}=\rho _{1}\,e^{\mathrm{i} \,S_{1}}\ ,\ \rho _{1}\geq 0 \;,
\\
\ \Phi _{3} &=&\rho _{2}\cos S_{2}\ ,\ \Phi _{4}=\rho _{2}\sin S_{2}
\
\Leftrightarrow \ \Phi _{3}=\RE \Psi _{2}\ ,\ \Phi _{4}=\IM \Psi _{2}
\ ,\ \
\Psi _{2}=\rho _{2}\,e^{\mathrm{i} \,S_{2}}\ ,\ \rho _{2}\geq 0 \;,
\end{eqnarray*}%
leading to the system in Eqs.~(\ref{mixFinal1}) and (\ref{mixFinal2}).

\subsection{On the relations with SUSY}
\label{sec4.2}

The important role played by the superpotentials in the present formalism
suggests a close relation with SUSY. In particular, one may suspect that
all the BPS systems constructed here correspond to some SUSY transformation
of suitable Fermionic fields in such a way that the present BPS equations
are the requirement to preserve a suitable amount of SUSY (see
\cite{SUSY} and references therein). However, we will now show that this
is not the case and that there are examples of spinor GPE that are not
supersymmetric (at least, not in the usual sense) and that (despite the
lack of SUSY) possess an associated BPS system. Let us consider Eqs.~(\ref{mix5})--(\ref{mix8}).
It is easy to see that this system of coupled partial differential equations
does not possess (generically) a Hamiltonian. In particular, let us analyze
the following system of coupled second-order PDEs for the four unknown
functions $\Phi _{j}\left ( x,y\right ) $ (we will consider the stationary
case for simplicity):
%
\begin{eqnarray}
\triangle \Phi _{1} &=&V_{1}\left ( \Phi _{1},\Phi _{2},\Phi _{3},
\Phi _{4}\right ) \ ,
\label{ex1}
\\
\triangle \Phi _{2} &=&V_{2}\left ( \Phi _{1},\Phi _{2},\Phi _{3},
\Phi _{4}\right ) \ ,\
\label{ex2}
\\
\triangle \Phi _{3} &=&V_{3}\left ( \Phi _{1},\Phi _{2},\Phi _{3},
\Phi _{4}\right ) \ ,
\label{ex3}
\\
\triangle \Phi _{4} &=&V_{4}\left ( \Phi _{1},\Phi _{2},\Phi _{3},
\Phi _{4}\right ) \ .
\label{ex4}
\end{eqnarray}%
Such a system possesses a Hamiltonian formulation when%
%
\begin{equation}
V_{j}=\frac{\partial P}{\partial \Phi _{j}}\quad ,\quad j\in \{1,2,3,4
\}\;.
\label{ham0}
\end{equation}%
In such a case $P$ is the self-interaction potential of the theory and
it is a trivial exercise to write down the corresponding Hamiltonian. On
the other hand, the right-hand sides
$V_{j}\left ( \Phi _{1},\Phi _{2},\Phi _{3},\Phi _{4}\right ) $ of Eqs.~(\ref{mix5})--(\ref{mix8}) read
%
\begin{eqnarray}
V_{1}\left ( \Phi _{1},\Phi _{2},\Phi _{3},\Phi _{4}\right ) &=&
\frac{1}{2}%
\frac{\partial }{\partial \Phi _{1}}\left ( A^{2}+B^{2}\right ) +C
\frac{%
\partial A}{\partial \Phi _{3}}+D
\frac{\partial B}{\partial \Phi _{3}}\ ,
\label{ham1}
\\
V_{2}\left ( \Phi _{1},\Phi _{2},\Phi _{3},\Phi _{4}\right ) &=&
\frac{1}{2}%
\frac{\partial }{\partial \Phi _{2}}\left ( A^{2}+B^{2}\right ) +C
\frac{%
\partial A}{\partial \Phi _{4}}+D
\frac{\partial B}{\partial \Phi _{4}}\ ,
\label{ham2}
\\
V_{3}\left ( \Phi _{1},\Phi _{2},\Phi _{3},\Phi _{4}\right ) &=&
\frac{1}{2}%
\frac{\partial }{\partial \Phi _{3}}\left ( C^{2}+D^{2}\right ) +A
\frac{%
\partial C}{\partial \Phi _{1}}+B
\frac{\partial D}{\partial \Phi _{1}}\ ,
\label{ham3}
\\
V_{4}\left ( \Phi _{1},\Phi _{2},\Phi _{3},\Phi _{4}\right ) &=&
\frac{1}{2}%
\frac{\partial }{\partial \Phi _{4}}\left ( C^{2}+D^{2}\right ) +A
\frac{%
\partial C}{\partial \Phi _{2}}+B
\frac{\partial D}{\partial \Phi _{2}}\ .
\label{ham4}
\end{eqnarray}%
It is clear that, in the case of Eqs.~(\ref{mix5})--(\ref{mix8}), the condition
in Eq.~(\ref{ham0}) is not satisfied (in the particular case of Eqs.~(\ref{sofe1})--(\ref{sofe4})
the condition in Eq.~(\ref{ham0}) is explicitly violated). It may still
be possible that for suitable choices of the superpotentials, the right-hand sides
$V_{j}\left ( \Phi _{1},\Phi _{2},\Phi _{3},\Phi _{4}\right ) $ of Eqs.~(\ref{mix5})--(\ref{mix8}) could satisfy Eq.~(\ref{ham0}), but the main
point here is that, generically, this is not the case.

Hence, if a system of coupled partial differential equations does not possess
a Hamiltonian, it cannot be supersymmetric (at least, not in the usual
sense) since one of the key properties of SUSY is that the Hamiltonian
is expressed in terms of the anti-commutators of the SUSY generators. Of
course, without Hamiltonian, such a property does not hold.

In conclusion, we have constructed infinitely many examples of coupled nonlinear
partial differential equations (some of which are relevant in the analysis
of relativistic field theories, of the multi-component GPE and so on) which
are not supersymmetric (in the usual sense) and which do possess a BPS
system nevertheless.

\subsection{General considerations for quantum mixtures}
\label{sec4.3}

The main goal of the present subsection is to disclose the power of the
present technique. In the case of quantum mixture, the moral of the story
is that once one has the GPE system of interest in a form similar to Eqs.~(\ref{mixture1}) and (\ref{mixture2}), then (to construct the sought BPS
system for the mixture), one has to choose the two superpotentials in Eqs.~(\ref{superp1}), (\ref{superp2}) and (\ref{superp3}) in such a way that
the right-hand sides of Eqs.~(\ref{mix5})--(\ref{mix8}) describes the typical
interactions of quantum mixtures (similar to the right-hand sides of Eqs.~(\ref{mixture1}) and (\ref{mixture2}).

For instance, a choice of superpotential which leads to natural cubic interaction
terms in the GPE for the mixture is%
%
\begin{eqnarray}
W_{1} &=&g_{1}\left ( Z_{1}\right ) ^{2}+g_{12}Z_{1}Z_{2}+g_{2}\left (
Z_{2}\right ) ^{2},\
\label{mixturesuperpotential}
\\
W_{2} &=&g_{3}\left ( Z_{1}\right ) ^{2}+g_{34}Z_{1}Z_{2}+g_{4}\left (
Z_{2}\right ) ^{2}\ ,
\notag
\end{eqnarray}%
where the parameters ($g_{1}$, $g_{2}$, $g_{3}$, $g_{4}$, $g_{12}$,
$g_{34}$%
) are related to the coupling constants appearing in the second-order GPE
for the mixture. With the above choice for the superpotential one obtains
\begin{eqnarray*}
A &=&\RE \left [ g_{1}\left ( \Phi _{1}+i\Phi _{2}\right ) ^{2}+g_{12}
\left ( \Phi _{1}+i\Phi _{2}\right ) \left ( \Phi _{3}+i\Phi _{4}
\right ) +g_{2}\left ( \Phi _{3}+i\Phi _{4}\right ) ^{2}\right ] \ ,
\ \
\\
B &=&\IM \left [ g_{1}\left ( \Phi _{1}+i\Phi _{2}\right ) ^{2}+g_{12}
\left ( \Phi _{1}+i\Phi _{2}\right ) \left ( \Phi _{3}+i\Phi _{4}
\right ) +g_{2}\left ( \Phi _{3}+i\Phi _{4}\right ) ^{2}\right ] \ ,
\\
\ \ C &=&\RE \left [ g_{3}\left ( \Phi _{1}+i\Phi _{2}\right ) ^{2}+g_{34}
\left ( \Phi _{1}+i\Phi _{2}\right ) \left ( \Phi _{3}+i\Phi _{4}
\right ) +g_{4}\left ( \Phi _{3}+i\Phi _{4}\right ) ^{2}\right ] \ ,
\ \
\\
D &=&\IM \left [ g_{3}\left ( \Phi _{1}+i\Phi _{2}\right ) ^{2}+g_{34}
\left ( \Phi _{1}+i\Phi _{2}\right ) \left ( \Phi _{3}+i\Phi _{4}
\right ) +g_{4}\left ( \Phi _{3}+i\Phi _{4}\right ) ^{2}\right ] \ .
\end{eqnarray*}
From the above expressions, one gets%
\begin{eqnarray*}
A &=&g_{1}\left ( \Phi _{1}^{2}-\Phi _{2}^{2}\right ) +g_{12}\left (
\Phi _{1}\Phi _{3}-\Phi _{2}\Phi _{4}\right ) +g_{2}\left ( \Phi _{3}^{2}-
\Phi _{4}^{2}\right ) \ ,\
\\
\ B &=&2g_{1}\Phi _{1}\Phi _{2}+g_{12}\left ( \Phi _{1}\Phi _{4}+
\Phi _{2}\Phi _{3}\right ) +2g_{2}\Phi _{3}\Phi _{4}\ ,
\\
C &=&g_{3}\left ( \Phi _{1}^{2}-\Phi _{2}^{2}\right ) +g_{34}\left (
\Phi _{1}\Phi _{3}-\Phi _{2}\Phi _{4}\right ) +g_{4}\left ( \Phi _{3}^{2}-
\Phi _{4}^{2}\right ) \ \ ,\ \
\\
D &=&2g_{3}\Phi _{1}\Phi _{2}+g_{34}\left ( \Phi _{1}\Phi _{4}+\Phi _{2}
\Phi _{3}\right ) +2g_{4}\Phi _{3}\Phi _{4}\ .
\end{eqnarray*}%
Plenty of spinor GPE systems have been shown to possess first-order BPS
systems. We will return to the present formalism's applications to spinor
Bose condensates in a future publication.

\section{Final remarks}
\label{sec-5}

In the present manuscript, it has been shown that a large class of generalized
GPE equations (with, for instance, the non-linear interaction terms
$|\Psi |^{5}$ and $|\Psi |^{6}$) do possess suitable BPS bounds on the
energy which can be saturated if the corresponding first-order BPS system
is satisfied (moreover, the static solutions of the BPS systems are also
solutions of the second-order field equations). This is a remarkable achievement
since BPS systems carry very powerful tools that are not available for
the usual second-order GPE. The simplest topologically non-trivial solutions
of the BPS represent configurations with fractional vorticity (the corresponding
fraction is related to the degree of non-linearity of the self-interaction
potential). However, more general multi-vortex-like configurations can
also be found (but a numerical approach is necessary). Moreover, this approach
can also be extended to coupled GPEs describing quantum mixtures, and thanks
to the algebraic relation between the velocity and amplitude, one can visualize
this relation directly through plots and solve numerically more easily.
We plan to come back to this in a forthcoming work.

We have also discussed the relationship between present formalism and SUSY.
Despite the many standard features, we have constructed examples of spinor
GPE that are not supersymmetric (at least, not in the standard sense) and
that, in fact, possess a BPS system. This formalism allows the powerful
analytic tools of BPS solitons to be applied to the topologically non-trivial
configurations of spinor GPE (usually analyzed only numerically). We will
come back to this intriguing topic shortly.

Regarding the possible experimental settings where the predictions of these
techniques could be tested, there are (at least) two natural candidates:
the (fractional) quantum Hall effect and the supersolids. Firstly, as far
as the fractional Hall effect is concerned, it would be very interesting
to connect the fractional statistics with the fractional vorticity, which
appears naturally in our framework. Secondly, in the case of supersolids,
the experimental techniques will soon become precise enough to test the
appearance of configurations with fractional vorticity. We hope to come
back on this interesting issue in a future publication.

\section*{Acknowledgements}

The authors would like to thank the referees for their useful comments and
suggestions. F. C. has been funded by FONDECYT Grant No. 1240048. P.~P.
gladly acknowledge support from Charles University Research Center (UNCE
24/SCI/016).

\begin{appendices}

\section{Solutions of BPS equations as solutions of GPE for general $Z^{n}$}
\label{appendix_BPS_as_GPE}

In this appendix, we show explicitly that a superpotential
$W=\frac{\kappa}{\sqrt{n}}\,Z^{n}$ where
$Z=\Phi _{1} + \mathrm{i}\,\Phi _{2}$ fulfills the GPE
\begin{eqnarray*}
\Delta \,\Phi _{1} &=& \kappa ^{2}\, (\Phi _{1}^{2} + \Phi _{2}^{2})^{n-1}
\,\Phi _{1} \;,
\\
\Delta \,\Phi _{1} &=& \kappa ^{2}\, (\Phi _{1}^{2} + \Phi _{2}^{2})^{n-1}
\,\Phi _{2} \;.
\end{eqnarray*}
We will write the two real fields $\Phi _{1}$ and $\Phi _{2}$, such that
$\Phi _{1}=\rho \,\cos \,S$ and $\Phi _{1}=\rho \,\sin \,S$. Then
%
\begin{eqnarray}
\frac{\partial \,\Phi _{1}}{\partial x} +
\frac{\partial \,\Phi _{2}}{\partial y} &=& \frac{\alpha}{\sqrt{n}}\,
\rho ^{n}\,\cos(n\,S)\;,
\label{BPS_1}
\\
\frac{\partial \,\Phi _{1}}{\partial y} -
\frac{\partial \,\Phi _{2}}{\partial x} &=& \frac{\alpha}{\sqrt{n}}\,
\rho ^{n}\,\sin(n\,S)\;,
\label{BPS_2}
\end{eqnarray}
where $n\in \mathbb{N}$, and $\alpha $ is a real constant. By deriving the first equation with respect to $x$ and the second with respect to
$y$ and by adding these results
\begin{eqnarray*}
\frac{\partial ^{2}\,\Phi _{1}}{\partial x^{2}} +
\frac{\partial ^{2}\,\Phi _{1}}{\partial y^{2}} &=& \alpha \,\sqrt{n}
\,\rho ^{n-1}\,\frac{\partial \,\rho}{\partial x}\,\cos(n\,S) -
\alpha \,\sqrt{n}\,\rho ^{n}\,\frac{\partial \,S}{\partial x}\,
\sin(n\,S)
\\
& & + \alpha \,\sqrt{n}\,\rho ^{n-1}\,
\frac{\partial \,\rho}{\partial y}\,\sin(n\,S) + \alpha \,\sqrt{n}\,
\rho ^{n}\,\frac{\partial \,S}{\partial y}\,\cos(n\,S) \;,
\end{eqnarray*}
It is useful to write
\begin{eqnarray*}
\frac{\partial \,\rho}{\partial x} &=&
\frac{\Phi _{1}\,\frac{\partial \,\Phi _{1}}{\partial \,x}+\Phi _{2}\,\frac{\partial \,\Phi _{2}}{\partial \,x}}{\rho}
\;, \quad \frac{\partial \,\rho}{\partial y} =
\frac{\Phi _{1}\,\frac{\partial \,\Phi _{1}}{\partial \,y}+\Phi _{2}\,\frac{\partial \,\Phi _{2}}{\partial \,y}}{\rho}
\;,
\\
\frac{\partial \,S}{\partial x} &=&
\frac{\Phi _{1}\,\frac{\partial \,\Phi _{2}}{\partial \,x}-\Phi _{2}\,\frac{\partial \,\Phi _{1}}{\partial \,x}}{\rho ^{2}}
\;, \quad \frac{\partial \,S}{\partial y} =
\frac{\Phi _{1}\,\frac{\partial \,\Phi _{2}}{\partial \,y}-\Phi _{2}\,\frac{\partial \,\Phi _{1}}{\partial \,y}}{\rho ^{2}}
\;.
\end{eqnarray*}
By using these expressions and rearranging terms, we get
\begin{eqnarray*}
\Delta \,\Phi _{1} &=& \kappa \,\sqrt{n}\,\rho ^{n-2}\left [\Phi _{1}
\,\cos(n\,S) + \Phi _{2}\,\sin(n\,S)\right ]\,\left (
\frac{\partial \,\Phi _{1}}{\partial \,x} +
\frac{\partial \,\Phi _{2}}{\partial \,y}\right )
\\
&&{}+ \kappa \,\sqrt{n}\,\rho ^{n-2}\left [\Phi _{1}\,\sin(n\,S) -
\Phi _{2}\,\cos(n\,S)\right ]\,\left (
\frac{\partial \,\Phi _{1}}{\partial \,y} -
\frac{\partial \,\Phi _{2}}{\partial \,x}\right ) \;,
\end{eqnarray*}
and this suggests us to reusing (\ref{BPS_1})--(\ref{BPS_2}), to obtain
\begin{eqnarray*}
\Delta \,\Phi _{1} &=& \kappa ^{2}\,\rho ^{2n-2}\left [\Phi _{1}\,
\cos(n\,S) + \Phi _{2}\,\sin(n\,S)\right ]\,\cos(n\,S)
\\
&&{}+ \kappa ^{2}\,\rho ^{2n-2}\left [\Phi _{1}\,\sin(n\,S) - \Phi _{2}
\,\cos(n\,S)\right ]\,\sin(n\,S) \;,
\end{eqnarray*}
which, after some algebra and the identity
$\cos ^{2}(n\,S) + \sin ^{2}(n\,S)=1$, we get
\begin{equation*}
\Delta \,\Phi _{1} = \kappa ^{2}\,\rho ^{2n-2}\,\Phi _{1} = \kappa ^{2}(
\Phi _{1}^{2} + \Phi _{2}^{2})^{n-1}\,\Phi _{1}\;.
\end{equation*}

Now, by deriving the first equation with respect to $y$ and the second with respect to $x$ and by subtracting these results,
\begin{equation*}
\frac{\partial ^{2}\,\Phi _{2}}{\partial x^{2}} +
\frac{\partial ^{2}\,\Phi _{2}}{\partial y^{2}} = \kappa \,\sqrt{n}\,
\rho ^{n-1}\,\frac{\partial \,\rho}{\partial y}\,\cos(n\,S) - \kappa
\,\sqrt{n}\,\rho ^{n}\,\frac{\partial \,S}{\partial y}\,\sin(n\,S) -
\kappa \,\sqrt{n}\,\rho ^{n-1}\,\frac{\partial \,\rho}{\partial x}\,
\sin(n\,S) - \kappa \,\sqrt{n}\,\rho ^{n}\,
\frac{\partial \,S}{\partial x}\,\cos(n\,S) \;,
\end{equation*}
and, after the same manipulations,
\begin{equation*}
\Delta \,\Phi _{2} = \kappa ^{2}\,\rho ^{2n-2}\,\Phi _{2}=\kappa ^{2}(
\Phi _{1}^{2} + \Phi _{2}^{2})^{n-1}\,\Phi _{2} \;.
\end{equation*}

Even more, we can write the right-hand side of (\ref{BPS_1})--(\ref{BPS_2})
in terms of $\Phi _{1}$ and $\Phi _{2}$ by using the $\cos(n\,S)$ and
$\sin(n\,S)$. Indeed, as
\begin{eqnarray*}
\rho ^{n}\cos(n\,S) &=& \rho ^{n} \sum _{r=0}^{[n/2]}\,(-1)^{r}
\binom{n}{2r}\,\cos ^{n-2r}(S)\,\sin ^{2r}(S) \;,
\\
&=& \sum _{r=0}^{[n/2]}\,(-1)^{r} \binom{n}{2r}\,\rho ^{n-2r}\,\cos ^{n-2r}(S)
\,\rho ^{2r}\sin ^{2r}(S)
\\
&=& \sum _{r=0}^{[n/2]}\,(-1)^{r} \binom{n}{2r}\,\Phi _{1}^{n-2r}\,
\Phi _{2}^{2r}\;,
\\
\rho ^{n}\sin(n\,S) &=& \rho ^{n} \sum _{r=0}^{[(n-1)/2]}\,(-1)^{r}
\binom{n}{2r+1}\,\cos ^{n-2r-1}(S)\,\sin ^{2r+1}(S) \;,
\\
&=& \sum _{r=0}^{[n/2]}\,(-1)^{r} \binom{n}{2r+1}\,\rho ^{n-2r-1}\,
\cos ^{n-2r-1}(S)\,\rho ^{2r+1}\sin ^{2r+1}(S)
\\
&=& \sum _{r=0}^{[(n-1)/2]}\,(-1)^{r} \binom{n}{2r+1}\,\Phi _{1}^{n-2r-1}
\,\Phi _{2}^{2r+1}\;.
\end{eqnarray*}
Therefore, a solution of the system,
%
\begin{eqnarray}
\frac{\partial \,\Phi _{1}}{\partial x} +
\frac{\partial \,\Phi _{2}}{\partial y} &=& \frac{\kappa}{\sqrt{n}}\,
\sum _{r=0}^{[n/2]}\,(-1)^{r} \binom{n}{2r}\,\Phi _{1}^{n-2r}\,\Phi _{2}^{2r}
\;,
\label{BPS_gen_1}
\\
\frac{\partial \,\Phi _{1}}{\partial y} -
\frac{\partial \,\Phi _{2}}{\partial x} &=& \frac{\kappa}{\sqrt{n}}\,
\rho ^{n}\,\sum _{r=0}^{[(n-1)/2]}\,(-1)^{r} \binom{n}{2r+1}\,\Phi _{1}^{n-2r-1}
\,\Phi _{2}^{2r+1}\;,
\label{BPS_gen_2}
\end{eqnarray}
is a solution of
%
\begin{eqnarray}
\Delta \,\Phi _{1} &=& \kappa ^{2}(\Phi _{1}^{2} + \Phi _{2}^{2})^{n-1}
\,\Phi _{1} \;,
\label{eq91}
\\
\Delta \,\Phi _{2} &=& \kappa ^{2}(\Phi _{1}^{2} + \Phi _{2}^{2})^{n-1}
\,\Phi _{2} \;.
\label{eq92}
\end{eqnarray}

\section{Fractional vortex solutions}
\label{appendix_fractional_vortex_solution}

It is more convenient to use polar coordinates,
%
\begin{eqnarray}
\label{polar_coordinates}
r&=&\sqrt{x^{2}+y^{2}} \;,
\nonumber
\\
\theta &=&\arctan(\frac{y}{x}) \;.
\end{eqnarray}
Using the change of variables formalae, we get
\begin{eqnarray*}
\frac{\partial \,\Phi _{1}}{\partial x} &=&
\frac{\partial \,r}{\partial x}
\frac{\partial \,\Phi _{1}}{\partial r} +
\frac{\partial \,\theta}{\partial x}
\frac{\partial \,\Phi _{1}}{\partial \theta} =
\frac{\partial \,\rho}{\partial r}\,\cos \theta \,\cos \,S - \rho \,
\frac{\partial \,S}{\partial r}\,\cos \theta \,\sin \,S -
\frac{\partial \,\rho}{\partial \theta}\,
\frac{\sin \theta \,\cos \,S}{r} + \rho
\frac{\partial \,S}{\partial \theta}\,
\frac{\sin \theta \,\sin \,S}{r}
\\
\frac{\partial \,\Phi _{2}}{\partial y} &=&
\frac{\partial \,r}{\partial y}
\frac{\partial \,\Phi _{2}}{\partial r} +
\frac{\partial \,\theta}{\partial y}
\frac{\partial \,\Phi _{2}}{\partial \theta} =
\frac{\partial \,\rho}{\partial r}\,\sin \theta \,\sin \,S + \rho \,
\frac{\partial \,S}{\partial r}\,\sin \theta \,\cos \,S +
\frac{\partial \,\rho}{\partial \theta}\,
\frac{\cos \theta \,\sin \,S}{r} + \rho
\frac{\partial \,S}{\partial \theta}\,
\frac{\cos \theta \,\cos \,S}{r}
\\
\frac{\partial \,\Phi _{1}}{\partial y} &=&
\frac{\partial \,r}{\partial y}
\frac{\partial \,\Phi _{1}}{\partial r} +
\frac{\partial \,\theta}{\partial y}
\frac{\partial \,\Phi _{1}}{\partial \theta} =
\frac{\partial \,\rho}{\partial r}\,\sin \theta \,\cos \,S - \rho \,
\frac{\partial \,S}{\partial r}\,\sin \theta \,\sin \,S +
\frac{\partial \,\rho}{\partial \theta}\,
\frac{\cos \theta \,\cos \,S}{r} - \rho
\frac{\partial \,S}{\partial \theta}\,
\frac{\cos \theta \,\sin \,S}{r}
\\
\frac{\partial \,\Phi _{2}}{\partial x} &=&
\frac{\partial \,r}{\partial x}
\frac{\partial \,\Phi _{2}}{\partial r} +
\frac{\partial \,\theta}{\partial x}
\frac{\partial \,\Phi _{2}}{\partial \theta} =
\frac{\partial \,\rho}{\partial r}\,\cos \theta \,\sin \,S + \rho \,
\frac{\partial \,S}{\partial r}\,\cos \theta \,\cos \,S -
\frac{\partial \,\rho}{\partial \theta}\,
\frac{\sin \theta \,\sin \,S}{r} - \rho
\frac{\partial \,S}{\partial \theta}\,
\frac{\sin \theta \,\cos \,S}{r} \;.
\end{eqnarray*}
Then, by substituting in (\ref{BPS_1}) and (\ref{BPS_2}),
\begin{eqnarray*}
\frac{\partial \,\rho}{\partial r}\left (\cos \theta \,\cos \,S +
\sin \theta \,\sin \,S\right ) &+& \rho
\frac{\partial \,S}{\partial r}\left (-\cos \theta \,\sin \,S+\sin
\theta \,\cos \,S\right ) + \frac{1}{r}
\frac{\partial \,\rho}{\partial \theta}\left (-\sin \theta \,\cos \,S+
\cos \theta \,\sin \,S\right ) +
\\
&&\frac{\rho}{r}\frac{\partial \,S}{\partial \theta}\left (\sin
\theta \,\sin \,S+\cos \theta \,\cos \,S\right ) =
\frac{\kappa}{\sqrt{n}}\rho ^{n}\cos(n\,S) \;,
\\
\frac{\partial \,\rho}{\partial r}\left (\sin \theta \,\cos \,S -
\cos \theta \,\sin \,S\right ) &+& \rho
\frac{\partial \,S}{\partial r}\left (-\sin \theta \,\sin \,S-\cos
\theta \,\cos \,S\right ) + \frac{1}{r}
\frac{\partial \,\rho}{\partial \theta}\left (\cos \theta \,\cos \,S+
\sin \theta \,\sin \,S\right ) +
\\
&&\frac{\rho}{r}\frac{\partial \,S}{\partial \theta}\left (-\cos
\theta \,\sin \,S+\sin \theta \,\cos \,S\right ) =
\frac{\kappa}{\sqrt{n}}\rho ^{n}\sin(n\,S)
\end{eqnarray*}
Therefore, by using the angle sum formulae for sines and cosines,
\begin{eqnarray*}
\frac{\partial \,\rho}{\partial r}\,\cos(\theta -S) + \rho
\frac{\partial \,S}{\partial r}\sin(\theta -S) - \frac{1}{r}
\frac{\partial \,\rho}{\partial \theta}\sin(\theta -S) +
\frac{\rho}{r}\frac{\partial \,S}{\partial \theta}\cos(\theta -S) =
\frac{\kappa}{\sqrt{n}}\rho ^{n}\cos(n\,S) \;,
\\
\frac{\partial \,\rho}{\partial r}\sin(\theta -S) - \rho
\frac{\partial \,S}{\partial r}\cos(\theta -S) + \frac{1}{r}
\frac{\partial \,\rho}{\partial \theta}\cos(\theta -S) +
\frac{\rho}{r}\frac{\partial \,S}{\partial \theta}\sin(\theta -S) =
\frac{\kappa}{\sqrt{n}}\rho ^{n}\sin(n\,S) \;,
\end{eqnarray*}
and, by rearranging,
\begin{eqnarray*}
\left [\frac{\partial \,\rho}{\partial r}+\frac{\rho}{r}
\frac{\partial \,S}{\partial \theta}\right ]\,\cos(\theta -S) +
\left [\rho \frac{\partial \,S}{\partial r}- \frac{1}{r}
\frac{\partial \,\rho}{\partial \theta}\right ]\,\sin(\theta -S) =
\frac{\kappa}{\sqrt{n}}\rho ^{n}\cos(n\,S) \;,
\\
\left [\frac{\partial \,\rho}{\partial r}+\frac{\rho}{r}
\frac{\partial \,S}{\partial \theta}\right ]\,\sin(\theta -S) -
\left [\rho \frac{\partial \,S}{\partial r}- \frac{1}{r}
\frac{\partial \,\rho}{\partial \theta}\right ]\,\cos(\theta -S) =
\frac{\kappa}{\sqrt{n}}\rho ^{n}\sin(n\,S) \;,
\end{eqnarray*}
Let us define the functions
\begin{eqnarray*}
\alpha (r,\theta ) &\equiv & \frac{\partial \,\rho}{\partial r}+
\frac{\rho}{r}\frac{\partial \,S}{\partial \theta} \;,
\\
\beta (r,\theta ) &\equiv & \rho \frac{\partial \,S}{\partial r}-
\frac{1}{r}\frac{\partial \,\rho}{\partial \theta} \;,
\end{eqnarray*}
then,
\begin{eqnarray*}
\alpha \,\cos(\theta -S) + \beta \,\sin(\theta -S) =
\frac{\kappa}{\sqrt{n}}\rho ^{n}\cos(n\,S) \;,
\\
\alpha \,\sin(\theta -S) - \beta \,\cos(\theta -S) =
\frac{\kappa}{\sqrt{n}}\rho ^{n}\sin(n\,S) \;,
\end{eqnarray*}
By defining $\gamma =\arctan(\frac{\beta}{\alpha})$, and performing some algebraic manipulations, we get
\begin{eqnarray*}
\alpha ^{2} + \beta ^{2} = \frac{\kappa ^{2}}{n}\,\rho ^{2n} \;,
\\
\tan(\theta -S-\delta ) = \tan \,(n\,S) \;.
\end{eqnarray*}
The last equation is only possible if $\delta = \theta - (n+1)\,S$. Therefore,
\begin{eqnarray*}
\alpha =\frac{\kappa}{\sqrt{n}}\,\rho ^{n}\,\cos \,\gamma &=&
\frac{\kappa}{\sqrt{n}}\rho ^{n}\,\cos \,(\theta -(n+1)\,S) \;,
\\
\beta =\frac{\kappa}{\sqrt{n}}\,\rho ^{n}\,\sin \,\gamma &=&
\frac{\kappa}{\sqrt{n}}\rho ^{n}\,\sin \,(\theta -(n+1)\,S) \;,
\end{eqnarray*}
or,
\begin{eqnarray*}
\frac{\partial \,\rho}{\partial r}+\frac{\rho}{r}
\frac{\partial \,S}{\partial \theta} &=& \frac{\kappa}{\sqrt{n}}\rho ^{n}
\,\cos \,(\theta -(n+1)\,S) \;,
\\
\rho \frac{\partial \,S}{\partial r}- \frac{1}{r}
\frac{\partial \,\rho}{\partial \theta}&=& \frac{\kappa}{\sqrt{n}}
\rho ^{n}\,\sin \,(\theta -(n+1)\,S) \;,
\end{eqnarray*}
This equation system is very general.

Let us suppose $\frac{\partial \,\rho}{\partial \theta}=0$ and
$\frac{\partial \,S}{\partial r}=0$. Then, $S=\frac{\theta}{n+1}$ and
\begin{equation*}
\frac{\partial \,\rho}{\partial r}+\frac{\rho}{r\,(n+1)} =
\frac{\kappa}{\sqrt{n}}\rho ^{n}\;,
\end{equation*}
or,
\begin{equation*}
\frac{1}{\rho ^{n}}\,\frac{\partial \,\rho}{\partial r}+
\frac{1}{\rho ^{n-1}\,r\,(n+1)} = \frac{\kappa}{\sqrt{n}} \;,
\end{equation*}
where we take $u\equiv \frac{1}{\rho}$.

When $n\geq 2$, we can use the change of variable
\begin{equation*}
u(r)=\rho ^{1-n} \;,
\end{equation*}
to obtain,
\begin{equation*}
u'+\frac{(1-n)}{(n+1)}\,\frac{u}{r} = \frac{(1-n)\,\kappa}{\sqrt{n}}
\;,
\end{equation*}
This is satisfied if $u(r)=u_{H}(r)+u_{P}(r)$ where $u_{H}$ is a solution
of the homogeneous equation
%
\begin{equation}
u'+\frac{(1-n)}{(n+1)}\,\frac{u}{r}=0\;,
\label{eq94}
\end{equation}
whose solution is $u_{H}(r)=A\,r^{\frac{n-1}{n+1}}$, where $A$ is a constant.
A particular solution $u_{P}(r)=B\,r^{q}$
%
\begin{equation}
B\,q\,r^{q-1}+\frac{(1-n)\,B}{(n+1)}r^{q-1} =
\frac{(1-n)\,\kappa}{\sqrt{n}} \quad \Rightarrow \quad q=1 \quad
\text{and} \quad B=\frac{(1-n^{2})\,\kappa}{2\sqrt{n}}\;.
\label{eq95}
\end{equation}
Then, $u_{P}(r)=\frac{(1-n^{2})\,\kappa}{2\sqrt{n}}\,r$, and
$u(r)=A\,r^{\frac{n-1}{n+1}} + \frac{(1-n^{2})\,\kappa}{2\sqrt{n}}\,r$,
giving us,
%
\begin{equation}
\label{rho_ngeq2}
\rho (r)=(A\,r^{\frac{n-1}{n+1}} +
\frac{(1-n^{2})\,\kappa}{2\sqrt{n}}\,r)^{\frac{1}{1-n}} \;.
\end{equation}
Finally, we get the two fields
\begin{eqnarray*}
\Phi _{1}(r,\theta ) &=& (A\,r^{\frac{n-1}{n+1}} +
\frac{(1-n^{2})\,\kappa}{2\sqrt{n}}\,r)^{\frac{1}{1-n}}\,
\cos(\frac{\theta}{n+1})\;,
\\
\Phi _{2}(r,\theta ) &=& (A\,r^{\frac{n-1}{n+1}} +
\frac{(1-n^{2})\,\kappa}{2\sqrt{n}}\,r)^{\frac{1}{1-n}}\,
\sin(\frac{\theta}{n+1}) \;.
\end{eqnarray*}
The results of the main text are obtained when $n=2$,
$n=\frac{5}{2}$ and $n=3$.

\appendix{Amplitude-phase relation for BPS solutions}
\label{appendix_amplitude_phase_relation}

Here, we will show that a solution of the BPS equation does not have a
totally independent amplitude $\rho $ and phase $S$.

Suppose $\Phi _{1}$ and $\Phi _{2}$ are two solutions of BPS equations
(\ref{BPS_1})--(\ref{BPS_2}), and we know the phase function
$S(r,\theta )$. By using some trigonometric formulae, we get
\begin{eqnarray*}
(\frac{\partial \,\rho}{\partial x} + \rho \,
\frac{\partial \,S}{\partial y}%
)\,\cos \,S + (\frac{\partial \,\rho}{\partial y} - \rho \,
\frac{\partial \,S}{%
\partial x})\,\sin \,S &=& \frac{\kappa}{\sqrt{n}}\,\rho ^{2}\,\cos
\,(n\,S) \;,
\\
(\frac{\partial \,\rho}{\partial y} - \rho \,
\frac{\partial \,S}{\partial x}%
)\,\cos \,S - (\frac{\partial \,\rho}{\partial x} + \rho \,
\frac{\partial \,S}{%
\partial y})\,\sin \,S &=& \frac{\kappa}{\sqrt{n}}\,\rho ^{2}\,\sin
\,(n\,S) \;.
\end{eqnarray*}
By changing the variable $u=\frac{1}{\rho}$, these equations are
\begin{eqnarray*}
(-\frac{\partial \,u}{\partial x} + u\,
\frac{\partial \,S}{\partial y}%
)\,\cos \,S + (-\frac{\partial \,u}{\partial y} - u\,
\frac{\partial \,S}{%
\partial x})\,\sin \,S &=& \frac{\kappa}{\sqrt{n}}\,\cos \,(n\,S) \;,
\\
(-\frac{\partial \,u}{\partial y} - u\,
\frac{\partial \,S}{\partial x}%
)\,\cos \,S - (-\frac{\partial \,u}{\partial x} + u\,
\frac{\partial \,S}{%
\partial y})\,\sin \,S &=& \frac{\kappa}{\sqrt{n}}\,\sin \,(n\,S) \;.
\end{eqnarray*}
As we know the function $S(x,y)$, these two equations must be satisfied
by the only unknown function $u(x,y)$, implying some constraints. Let us
write
%
\begin{eqnarray}
\label{general_system}
A_{1}\,\frac{\partial \,u}{\partial x} + B_{1}\,
\frac{\partial \,u}{\partial y} + C_{1}\,u&=& F_{1} \;,
\\
A_{2}\,\frac{\partial \,u}{\partial x} + B_{2}\,
\frac{\partial \,u}{\partial y} + C_{2}\,u&=& F_{2} \;,
\label{eq98}
\end{eqnarray}
where
\begin{eqnarray*}
A_{1} = -\cos \,S\;, \quad B_{1}=-\sin \,S\;, \quad C_{1}=
\frac{\partial \,S}{%
\partial y}\cos \,S - \frac{\partial \,S}{\partial x}\sin \,S\;, & & F_{1}
= \frac{\kappa}{\sqrt{n}}\,\cos \,(n\,S)\;,
\\
A_{2}=\sin \,S\;,\ \quad B_{2}=-\cos \,S\, \quad C_{2}= -
\frac{\partial \,S}{%
\partial x}\cos \,S - \frac{\partial \,S}{\partial y}\sin \,S\;, & & F_{2}
= \frac{\kappa}{\sqrt{n}}\,\sin \,(n\,S) \;,
\end{eqnarray*}
By dividing the first equation by $A_{1}$ and the second by $A_{2}$, we
get,
\begin{eqnarray*}
\frac{\partial \,u}{\partial x} + \tilde{B}_{1}\,
\frac{\partial \,u}{\partial y%
} + \tilde{C}_{1}\,u&=& \tilde{F}_{1} \;,
\\
\frac{\partial \,u}{\partial x} + \tilde{B}_{2}\,
\frac{\partial \,u}{\partial y%
} + \tilde{C}_{2}\,u&=& \tilde{F}_{2} \;,
\end{eqnarray*}
and, by substracting the first to the second,
\begin{equation*}
(\tilde{B}_{2}-\tilde{B}_{1})\,\frac{\partial \,u}{\partial y} + (
\tilde{C}%
_{2}-\tilde{C}_{1})\,u = \tilde{F}_{2} - \tilde{F}_{1} \;.
\end{equation*}
It can be shown
\begin{eqnarray*}
\tilde{B}_{2}-\tilde{B}_{1} &=& -\cot \,S - \tan \,S\;,
\\
\tilde{C}_{2}-\tilde{C}_{1} &=& -\frac{\partial \,S}{\partial x}%
\,(\tan \,S+\cot \,S)\;,
\\
\tilde{F}_{2}-\tilde{F}_{1} &=&\frac{\kappa}{\sqrt{n}}\,\kappa \left (
\frac{\cos \,(n\,S)%
}{\cos \,S} + \frac{\sin \,(n\,S)}{\sin \,S}\right )\;.
\end{eqnarray*}
Therefore,
%
\begin{equation}
\label{condition_u_y_multivortex}
\frac{\partial \,u}{\partial y} - A(x,y)\,u = B(x,y) \;.
\end{equation}
where we defined the functions
\begin{eqnarray*}
A(x,y)&\equiv &-\frac{\partial \,S}{\partial x}\;,
\\
B(x,y)&\equiv & -\frac{\kappa}{\sqrt{n}}\,\sin \,((n+1)\,S)\;.
\end{eqnarray*}
The formal solution for equation (\ref{condition_u_y_multivortex}) is
%
\begin{equation}
\label{general_solution_u_y}
u(x,y)=e^{\int_{y_{0}}^{y}\,A(x,y^{\prime })\,dy^{\prime }}\left[ \alpha(x)
+ \int_{y_{0}}^{y}\,e^{-\int_{y_{0}}^{y^{\prime }}\,A(x,y^{\prime \prime
})\,dy^{\prime \prime }}\,B(x,y^{\prime })\,dy^{\prime }\right] \;,
\end{equation}
where $\alpha $ is a function only on $x$.

Now, let us come back to equation (\ref{general_system}), but this time,
we divide the first equation by $B_{1}$ and the second by $B_{2}$,
\begin{eqnarray*}
\tilde{A}_{1}\,\frac{\partial \,u}{\partial x} +
\frac{\partial \,u}{\partial y%
} + \tilde{\tilde{C}}_{1}\,u&=& \tilde{\tilde{F}}_{1} \;,
\\
\tilde{A}_{2}\,\frac{\partial \,u}{\partial x} +
\frac{\partial \,u}{\partial y} + \tilde{\tilde{C}}_{2}\,u&=&
\tilde{\tilde{F}}_{2} \;,
\end{eqnarray*}
and, by subtracting the first to the second,
\begin{equation*}
(\tilde{A}_{2}-\tilde{A}_{1})\,\frac{\partial \,u}{\partial x} + (
\tilde{%
\tilde{C}}_{2}-\tilde{\tilde{C}}_{1})\,u = \tilde{\tilde{F}}_{2} -
\tilde{%
\tilde{F}}_{1} \;.
\end{equation*}

It can be shown
\begin{eqnarray*}
\tilde{A}_{2}-\tilde{A}_{1} &=& -\cot \,S - \tan \,S\;,
\\
\tilde{\tilde{C}}_{2}-\tilde{\tilde{C}}_{1} &=&
\frac{\partial \,S}{\partial y%
} \left (\cot \,S + \tan \,S\right )\;,
\\
\tilde{\tilde{F}}_{2}-\tilde{\tilde{F}}_{1} &=&
\frac{\kappa}{\sqrt{n}}\left (\frac{%
\cos \,(n\,S)}{\sin \,S} - \frac{\sin \,(n\,S)}{\cos \,S}\right )\;.
\end{eqnarray*}
Therefore,
%
\begin{equation}
\label{condition_u_x_multivortex}
\frac{\partial \,u}{\partial x} - C(x,y)\,u = D(x,y) \;.
\end{equation}
where we defined the functions
\begin{eqnarray*}
C(x,y)&\equiv &\frac{\partial \,S}{\partial y}\;,
\\
D(x,y)&\equiv &-\frac{\kappa}{\sqrt{n}}\,\cos \,((n+1)\,S)\;.
\end{eqnarray*}
The formal solution for equation (\ref{condition_u_x_multivortex}) is
%
\begin{equation}
\label{general_solution_u_x}
u(x,y)=e^{\int_{x_{0}}^{x}\,C(x^{\prime },y)\,dx^{\prime }} \left[\beta(y) +
\int_{x_{0}}^{x}\,e^{-\int_{x_{0}}^{x^{\prime }}\,C(x^{\prime \prime
},y)\,dx^{\prime \prime }}\,D(x^{\prime },y)\,dx^{\prime }\right] \;,
\end{equation}
where $\beta $ is a function only on $y$.

Thus, functions (\ref{general_solution_u_y}) and (\ref{general_solution_u_x})
should be the same.

We can derivate the equation (\ref{condition_u_y_multivortex}) with respect
to $x$ and the equation (\ref{condition_u_x_multivortex}) with respect
to $y$ and subtracting,
\begin{eqnarray*}
0 = \frac{\partial u}{\partial y \, \partial x} -
\frac{\partial u}{\partial x \, \partial y} &=&
\frac{\partial u}{\partial x}A + u\,\frac{\partial A}{%
\partial x} + \frac{\partial B}{\partial x} -
\frac{\partial u}{\partial y}C - u\,\frac{\partial C}{\partial y} -
\frac{\partial D}{\partial y}
\\
&=& u\left (\frac{\partial A}{\partial x} -
\frac{\partial C}{\partial y}%
\right ) + AD-BC + \frac{\partial B}{\partial x} -
\frac{\partial D}{\partial y} \;.
\end{eqnarray*}
where in the last equality, we again used equations (\ref{condition_u_y_multivortex})
and (\ref{condition_u_x_multivortex}). That means,
\begin{equation*}
0= -u\left (\frac{\partial ^{2} S}{\partial x^{2}} +
\frac{\partial ^{2} S}{%
\partial y^{2}}\right )- \frac{\sqrt{n}\,\kappa}{(n+1)}\,\left (
\frac{%
\partial \,\cos ((n+1)\,S)}{\partial y}-
\frac{\partial \,\sin ((n+1)\,S)}{\partial x}\right ) \;,
\end{equation*}
One can go to the polar coordinates $(r,\theta )$ to get
%
\begin{equation}
\label{condition_S}
\Delta \,S = \sqrt{n}\,\kappa \,\rho ^{n-1}\,\left (
\cos ((n+1)\,S -
\theta )\,\frac{\partial \,S}{\partial r}+
\frac{\sin ((n+1)\,S - \theta )}{r}\,
\frac{\partial \,S}{\partial \theta}\right ) \;.
\end{equation}
One can check that for $n=3$ we get exactly the relation (\ref{amp_pha_rel_n3}).

\end{appendices}

\bibliographystyle{utphys}
\bibliography{ComplexGPE_paper_biblio}

\end{document}